\documentclass{aa}  

\usepackage{longtable}
\usepackage{graphicx}
\usepackage{color}
\usepackage{amsmath,amssymb}
\usepackage{url}
\usepackage{mathptmx}
 \newcommand\gsim{\lower.6ex\hbox{$\buildrel>\over\sim$}}
\newcommand\lsim{\lower.6ex\hbox{$\buildrel<\over\sim$}}
\usepackage{txfonts}
\begin{document}

   \title{The hunt for extraterrestrial high-energy neutrino counterparts}
\titlerunning{The Wild Hunt}
\authorrunning{Liodakis et al.}

\author{I. Liodakis\inst{\ref{inst1}}\thanks{yannis.liodakis@utu.fi} \and T. Hovatta\inst{\ref{inst1},}\inst{\ref{inst2}} \and V. Pavlidou\inst{\ref{inst4},}\inst{\ref{inst7}} \and A.C.S. Readhead\inst{\ref{inst4},}\inst{\ref{inst9}} \and R. D. Blandford\inst{\ref{inst3}} \and S. Kiehlmann\inst{\ref{inst4},}\inst{\ref{inst5}} \and E. Lindfors\inst{\ref{inst1}} \and W. Max-Moerbeck\inst{\ref{inst6}} \and T. J. Pearson\inst{\ref{inst9}} \and M. Petropoulou\inst{\ref{inst8}} }

\institute{Finnish Centre for Astronomy with ESO, University of Turku, Vesilinnantie 5, FI-20014, Finland\label{inst1}
\and  Aalto University Mets\"ahovi Radio Observatory, Mets\"ahovintie 114, 02540 Kylm\"al\"a, Finland \label{inst2}
\and Institute of Astrophysics, Foundation for Research and Technology-Hellas, GR-71110 Heraklion, Greece\label{inst4}
\and Department of Physics and Institute of Theoretical and Computational Physics, University of Crete, 71003 Heraklion, Greece\label{inst7}
\and Owens Valley Radio Observatory, California Institute of Technology,  Pasadena, CA 91125, USA\label{inst9}
\and Kavli Institute for Particle Astrophysics and Cosmology, Department of Physics, Stanford University, Stanford, CA 94305, USA \label{inst3}
\and Department of Physics, Univ. of Crete, GR-70013 Heraklion, Greece\label{inst5}
\and Departamento de Astronomi\'a, Universidad de Chile, Camino El Observatorio 1515, Las Condes, Santiago, Chile\label{inst6}
\and Department of Physics, National and Kapodistrian University of Athens, University Campus Zografos, GR 15783, Greece\label{inst8}
}


 
  \abstract{The origin of Petaelectronvolt (PeV) astrophysical neutrinos is fundamental to our understanding of the high-energy Universe. Apart from the technical challenges of operating detectors deep below ice, oceans, and lakes, the phenomenological challenges are even greater than those of gravitational waves; the sources are unknown, hard to predict, and we lack clear signatures. Neutrino astronomy therefore represents the greatest challenge faced by the astronomy and physics communities thus far. The possible neutrino sources range from accretion disks and tidal disruption events, to relativistic jets and galaxy clusters with blazar TXS~0506+056 the most compelling association thus far. Since that association, immense effort has been put into proving or disproving that jets are indeed neutrino emitters, but to no avail. By generating simulated neutrino counterpart samples, we explore the potential of detecting a significant correlation of neutrinos with jets from active galactic nuclei. We find that, given the existing challenges, even our best experiments could not have produced a $>3\sigma$ result. Larger programs over the next few years will be able to detect a significant correlation only if the brightest radio sources, rather than all jetted active galactic nuclei, are neutrino emitters. We discuss the necessary strategies required to steer future efforts into successful experiments.}


\maketitle

%

\section{Introduction} \label{sec:intro}

Throughout the history of modern astronomy, the introduction of new theories and new technologies has led to giant leaps in our understanding not only of astronomy, but also of physics and cosmology. The successive openings of the radio, X-ray, $\gamma-$ray, and most recently the gravitational wave windows, have revolutionized our view of the Universe.  The opening of the high-energy $\sim$PeV neutrino window is likely among the last observational frontiers. It is reasonable to expect that opening this window will lead to fundamental changes in our understanding,  as has happened in all previous cases. The start of operations of IceCube in 2010 and the detection of high-energy astrophysical neutrinos \citep{2013Sci...342E...1I,2014PhRvL.113j1101A} was a significant step towards this goal. Identifying sources of neutrinos is the next obvious, yet paramount, step forward.

In the one proven case of multi-messenger astrophysics, that is that of gravitational waves, each potential gravitational wave source has a distinctive gravitational wave signature. Therefore we know what we are looking at and what we are looking for. In the case of multi-messenger neutrino astrophysics there are no distinctive neutrino signatures. This is exacerbated by the fact that at high energies we may only receive one neutrino from each source, which is not sufficient to provide a neutrino signature. To make things worse, there is a huge background of atmospheric muons ($\sim 10^{11}\;{\rm yr}^{-1}$) and atmospheric neutrinos ($\sim 10^{5}\;{\rm yr}^{-1}$) against which the high-energy astrophysical neutrinos ($\sim 10\;{\rm yr}^{-1}$) have to be detected. Nevertheless, in 2017 IceCube observed a high-energy neutrino event designated as IceCube-170922A. The best-fit  reconstructed  direction  was at  0.1$^\circ$ from the  sky  position  of  the  blazar [an active galactic nucleus (AGN) with a jet pointed towards the observer] TXS~0506+056, which was flaring in radio, in X-rays and in GeV and TeV $\gamma$-rays \citep{2017ATel10791....1T,2017ATel10817....1M,2019MNRAS.483L..42K}. The positional and temporal coincidences were significant at the 3$\sigma$ level, suggesting the detection of the first counterpart to a high-energy extragalactic neutrino \citep{2018Sci...361.1378I}. There is one other radio bright jetted-AGN within 2 degrees of the IceCube position, and the number of sources is expected to scale as S$^{-3/2}$.

 In relativistic jets, there is now evidence that electromagnetic (EM) emission can originate in the outer surface of the jet, enhanced by interaction with the interstellar medium (ISM, \citealp{Kim2018,Janssen2021}). In this situation, one could expect entrainment of ISM material, and hence of hadrons, that would give rise to a hadronic emission channel in relativistic jets. Thus relativistic jets in AGN could produce neutrinos and it is concerning that this association remains in doubt after over a decade of observations by IceCube and the detection of dozens of high-energy astrophysical neutrinos. As confidence in the existence of astrophysical high-energy neutrinos grows through the increasingly sophisticated removal of the contaminating atmospheric muons and neutrinos, if no EM counterparts are found, the possibility of dark matter sources and of physics beyond the standard model gains strength. These are compelling arguments for mounting the most powerful attacks possible for the identification of the EM sources of astrophysical neutrinos. This paper has been written partly in order to draw attention to this strategic deficit in the pursuit of multi-messenger neutrino astrophysics.

Discussions of high-energy (TeV-PeV) neutrinos in relativistic jets have involved pion production through proton-proton \citep[e.g.,][]{Loeb2006} and proton-photon interactions \citep[e.g.,][]{Stecker1991,Mannheim1995,Halzen1997}.  This requires proton acceleration to energies typically an order of magnitude larger than the observed neutrino energies. Interestingly, these processes predict an initial flavor ratio of $\nu_e$ : $\nu_\mu$ : $\nu_\tau$ of 1 : 2 : 0 for $pp$ collisions that produce roughly equal number of $\pi +$ and $\pi -$.  For $p \gamma$ interactions close to the $\Delta$ resonance, the relative ratio can be 1:1:0 \citep{2005PhLB..621...18A}. The neutrinos then mix, in an energy-dependent manner, on the journey to Earth. The mixing ratios could be confidently studied by an expanded IceCube facility such as IceCube-Gen2 \citep{2020arXiv200804323T}.

 Efficient particle acceleration in blazar jets is quite natural. The most luminous flat spectrum radio quasars (FSRQ) near their source can generate up to $\sim1\,{\rm ZV}=10^{21}\,{\rm V}$ of electromotive force and, for this reason, quasar jets have commonly been suspected as the source of ultra high energy cosmic rays with energy up to $\sim10^{20}\,{\rm eV}$ (e.g., \citealp{Globus2017,Mbarek2021}). It is therefore quite reasonable that protons, given the longer cooling timescales than leptons, can be accelerated to high enough energies to account for the observed astrophysical neutrinos. However, the charged pions produced in these interactions also create pairs which will radiate $\gamma$-rays either through inverse-Compton scattering or synchrotron radiation. Neutral pions create $\gamma$-rays directly. Following the detection of very-high-energy $\gamma$-rays and the IceCube-170922A event from the direction of the jetted-AGN  TXS~0506+056, various theoretical models have been applied to explain the observed association of neutrinos and blazar flares \citep[see e.g.,][]{2018ApJ...863L..10A,2018ApJ...866..109S, murase18, keivani18, 2019ApJ...876..109Z, 2020ApJ...891..115P}  as well as neutrinos and steady emission processes of blazar jets \citep[see e.g.,][]{2018ApJ...865..124M, 2019ApJ...874L..29R}. Many of these papers show that it is not easy to explain the multi-messenger data of TXS~0506+056. First, as suggested for example by \cite{keivani18} and \cite{2019ApJ...876..109Z}, a physically consistent picture can only be achieved with a hybrid leptonic scenario, with $\gamma$-rays produced by external inverse-Compton processes and high-energy neutrinos produced by photopion production with the external radiation field. The best-fit models, however, typically produce neutrinos peaked at higher energy but with lower flux compared to observations. 
Alternatively, a hadronuclear interaction channel has also been investigated \citep[see e.g.,][]{2019PhRvD..99f3008L}. However, producing the observed neutrino flux through proton-proton interactions requires an extreme jet power for accelerating relativistic protons that significantly exceeds the Eddington luminosity \citep[see also,][]{Liodakis2020}.  In addition single-zone models, where $\gamma$-ray and neutrino emission are produced in the same location, can hardly reproduce the broadband spectral energy distribution and the temporal profiles  of TXS~0506+056 simultaneously. As a result, more complicated multi-zone models are likely to be needed \citep{Murase2018, 2019MNRAS.483L..12C}. \cite{2019ApJ...874L...9H} and \cite{2019ApJ...881...46R} showed that the absorption and interactions intrinsic to the source, and by the extragalactic background, may significantly alter the $\gamma-$ray spectrum.  Internal absorption may cause a lack of high-energy $\gamma-$ray and neutrino correlation in AGN \citep{2017ApJ...835...45A} and maybe hint towards a correlation with lower energy observations in the X-ray and radio bands. In summary, despite multi-wavelength follow-ups of the event and extensive theoretical modeling, the radiation mechanisms of TXS~0506+056 and the underlying magnetic field strength and configuration remain poorly understood. The theory of neutrino production in blazar jets is still largely unknown.

Following the identification of TXS~0506+056 the majority of efforts have focused in the $\gamma$-ray regime. However, this connection has proven complicated.  The regions that would host sufficiently dense external photon fields to produce high neutrino fluxes would also imply strong internal absorption of $\gamma$-rays \citep[see e.g.,][]{Murase2016,2019ApJ...881...46R}, and therefore no temporal coincidence would be expected. Indeed, no $\gamma$-ray flare was detected from TXS~0506+056 during the 2014--2015 neutrino burst \citep{2018Sci...361.1378I}. Recent modeling results also suggest that $\gamma$-ray emission could be heavily suppressed during neutrino flares \citep{Kun2021,Mastichiadis2020}.

Further complications in identifying neutrino counterparts arise from the possibility that not all blazar classes are equally ``good'' neutrino candidates \citep[e.g.,][]{PhysRevD.66.123003}. \citet{2016MNRAS.457.3582P} found the most significant connection with high synchrotron peaked (HSP) sources\footnote{Blazars are often classified using the peak frequency of the synchrotron emission as low-, intermediate-, and high-synchrotron-peak blazars ($\nu_{syn}<10^{14}~Hz$, $10^{14}<\nu_{syn}<10^{15}~Hz$, and $\nu_{syn}>10^{15}~Hz$, respectively \citealp{Abdo2010}).}. The most recent revision (with more data) suggested a connection with intermediate synchrotron peaked (ISP) and HSP sources \citep{2020MNRAS.497..865G}. On the other hand, \cite{2019ICRC...36..916H} found the most significant connection (though only at $1.9\sigma$ level) with low synchrotron peaked (LSP) and ISP BL Lac objects, and the connection to HSP had a much lower significance (only $0.5\sigma$). \cite{2020ApJ...894..101P} carried out a careful study of 3,388 jetted-AGN for which  there are very long baseline interferometry (VLBI) images at 8 GHz, and a sample of 56 IceCube high-energy neutrino events $E>$200 TeV, and found significant evidence for an association, with a chance probability of only 0.2\%. They also found highly probable associations with the jetted-AGN 3C 279, NRAO 530, PKS 1741-038, and OR 103.   In a separate study,  using IceCube track data for seven years and 712,830 detected events, and 3,411 jetted AGN with VLBI data, \cite{2021ApJ...908..157P} found evidence for association with a post trial $p-$value of $3 \times 10^{-3}$, corresponding to a $3\sigma$ significance for a normal distribution.  The combined significance of these two studies is 4.1$\sigma$. Shortly after, this result was challenged by \cite{Zhou2021} using the same sample.

In a recent paper, we investigated the connection between astrophysical high-energy neutrinos and jetted-AGN monitored by the Owens valley radio observatory (OVRO) and Mets\"ahovi radio observatory \cite[][hereafter H21]{2021A&A...650A..83H}. OVRO has been monitoring a large sample of jetted-AGN since 2008 with regular cadence of about 3 days. The initial sample consisted of 1157 sources (hereafter CGRaBS sample). Over the years more sources detected by the {\it Fermi} gamma-ray space telescope ({\it Fermi}) were added to the sample, which at the time of H21 numbered 1795 sources (hereafter full sample). We considered three different samples: the CGRaBS sample, the full sample, and a flux-density limited sample at 350 mJy \citep{Liodakis2017}. We then cross-matched the coordinates of the OVRO sources with the reconstructed position of high-energy neutrinos, and evaluated the radio activity state of the coincident sources. Using radio light curves from OVRO, we identified 20 candidate sources positionally associated with astrophysical neutrinos during periods of enhanced radio activity at the $\sim2\sigma$ level. Almost half the potentially associated sources (9/20) have not been detected by {\it Fermi}. This is interesting considering that a large fraction of the OVRO monitoring sample was included after the sources were detected by {\it Fermi}, implying that $\gamma$-ray activity does not play a major role in the neutrino association. Additionally, the fact that they have not been detected by {\it Fermi}, places strong constraints on the possible location and neutrino emission mechanism as the jet physical conditions have to be such that protons are energetic enough to produce neutrinos, but at the same time not produce any detectable $\gamma$-ray emission. Alternatively, neutrino production has to take place in regions where the opacity for $\gamma \gamma$ pair production is very high. 

Apart from the $13\pm5$ neutrino excess above the atmospheric background found in TXS~0506+056, thus far all of the suggested neutrino associations only have a single high-energy neutrino detection per source (either an AGN or a tidal disruption event [TDE]). In the claimed TDE association \citep{2021NatAs...5..510S} the neutrino lags the TDE by $\sim 120$ days, and the association has been challenged  \citep{Cendes2021}. The claimed $\sim 3\sigma$ AGN association has also been questioned \citep{2019Galax...7...20B,2019ApJ...881...46R,Kun2021}. Considering different neutrino datasets yields a $<3\sigma$ probability of association with TXS 0506+056 \citep{Icecube2021}. The a posteriori nature of these associations is not conducive to establish real connections with AGN, since a posteriori probabilities are notoriously difficult to calculate.
Two of the studies based on well-defined samples that took the a priori approach showed high-energy neutrino-AGN associations significant at the $2.9\sigma$ \citep{2020ApJ...894..101P} and $3.3\sigma$ \citep{Icecube2021} levels.

In this paper, we focus on the potential connection between blazars and astrophysical neutrinos. We lay out the necessary strategies that will allow us to prove or disprove AGN with jets as the sources of high-energy neutrinos and explore the shortcomings of our current efforts. In sections \ref{sec:strat} and \ref{sec:strat2} we lay out and discuss different strategies. In section \ref{sec:sims} we simulate neutrino counterpart samples and explore the detectability of a statistical significant correlation with jetted-AGN by current and future experiments. We draw conclusions in section \ref{sec:concl}.

\section{General strategies for identifying neutrino counterparts}\label{sec:strat}

Since we lack compelling proof that high-energy astrophysical neutrinos are associated with AGN, it is important to be as unbiased as possible in our approach. It would be just as important to disprove as to prove that there is such an association. With this in mind, we must consider how to optimize our chances.  This leads us to the following conclusions regarding the optimization of any potential experiment with that goal. Firstly, it should establish clear a priori hypotheses to test,  and not change the criteria once observing has begun. Clearly these hypotheses can be applied only to observations that commence after these hypotheses are laid down, that is only to data taken after the start of any experiment. In addition, sample selection should be unbiased and the sample should not try to mimic properties of existing candidates, but rather be motivated by theory. Secondly, Earth shielding against atmospheric muons is important in obtaining as clean a sample of astrophysical neutrinos as possible.  Therefore this experiment should concentrate primarily on neutrinos from the hemisphere opposite to that of the neutrino detector. Thirdly,  as discussed above, it is possible that not all types of blazars are good neutrino candidates, and some might be $\gamma-$ray dark. Sources that intrinsically do not produce neutrinos will dilute the sample. In this case, looking at the significance of individual sub-classes is preferable to the combined analysis. Therefore, any sample should include as many different classes of jetted-AGN as possible. Finally, the probability of finding an association with a flare is proportional to the observing time, thus at least a few years of monitoring are necessary. To the best of our knowledge only the current monitoring program by OVRO meets the above criteria.

\section{Positional association alone is futile}\label{sec:strat2}

The futility of spatial-only searches for neutrino-AGN associations becomes obvious  when considering that there are $\gsim 10^3$ AGN in each IceCube positional uncertainty area on the sky. We came to the same conclusion in H21. Here, we further demonstrate the need for timing information.

We consider below the statistical significance of associations using the sample of neutrinos from our work in H21 and a simple binomial calculation. Let us simplify the association of jetted-AGN and neutrinos into a ``dart throwing'' experiment. We have a number of $N_\mathrm{neut}$ neutrinos detected by IceCube, which have a certain positional uncertainty region around them. We then attempt to ``hit'' these neutrinos with our sample of $n$ jetted-AGN. If the jetted-AGN falls within the uncertainty ellipse of a neutrino, we call that a success. Then the probability of getting at least $k$ successes just by chance, that is the estimate of how unlikely $k$ successes are, can be obtained by the binomial cumulative distribution function
\begin{equation}\label{eq:binomial}
P(k~\mathrm{or~more}) = \sum\limits_{i=k}^{n}\binom{n}{i}p_\mathrm{chance}^i(1-p_\mathrm{chance})^{n-i},
\end{equation}
where $p_\mathrm{chance}$ is the probability for an individual jetted-AGN to fall within the uncertainty ellipses of the $N_{\rm neut}$ neutrino events on the null hypothesis that neutrino events arise from an isotropic population.

We can estimate $p_\mathrm{chance}$ by using the sample of $N_\mathrm{neut}=56$ high-energy neutrinos detected by IceCube in 2009-2019 used in H21. The mean uncertainty ellipse radius, including also a systematic uncertainty of $0.9^\circ$ added in quadrature, was found to be $1.5^\circ$, resulting in an $\epsilon = 7.1~{\rm deg}^2$ uncertainty region on the sky around each neutrino. Because OVRO only observes sources at declination $\delta>-20^\circ$, we consider an area of $A = 27681~{\rm deg}^2$ on the sky.

Now the chance probability for a single jetted-AGN to fall within the uncertainty ellipses of the neutrinos is 
\begin{equation}
    p_\mathrm{chance} = \frac{N_\mathrm{neut} \epsilon}{A} = 0.014.
\end{equation}

To demonstrate the calculation of the binomial probability, we used the sample of 1157 CGRaBS sources from H21, which is our number of trials $n$. We find that 17 of these fall within the error ellipse of the $N_\mathrm{neut}=56$ neutrinos, meaning that our number of successes is $k=17$. Using Eq.~\ref{eq:binomial} we find that the probability to have at least this many associations by chance is $P (k \geq 17) = 0.49$, hence we cannot claim a connection based on positional association alone. 

We now repeat the exercise for 5000 sources, assuming the same 56 neutrino events, as this is close to the expected number of events in the next 5 years. The number of sources is motivated by OVRO's current 5-year monitoring program dedicated to the association of neutrinos with jetted-AGN. In this case, the chance probability for a single event is the same as before, $p_\mathrm{chance} = 0.014$. However, now we have $n=5000$ jetted-AGN, and the number of matches increases to $k=64$. We note that this is actually higher than the number of observed neutrinos $N_\mathrm{neut}=56$ because multiple sources can fall within the same uncertainty ellipse. The binomial probability will be $P (k \geq 64) = 0.84$, which is completely uninformative.

\section{Detecting a correlation between AGN and neutrinos}\label{sec:sims}

\begin{figure}
\centering
\resizebox{\hsize}{!} {\includegraphics[width=\hsize]{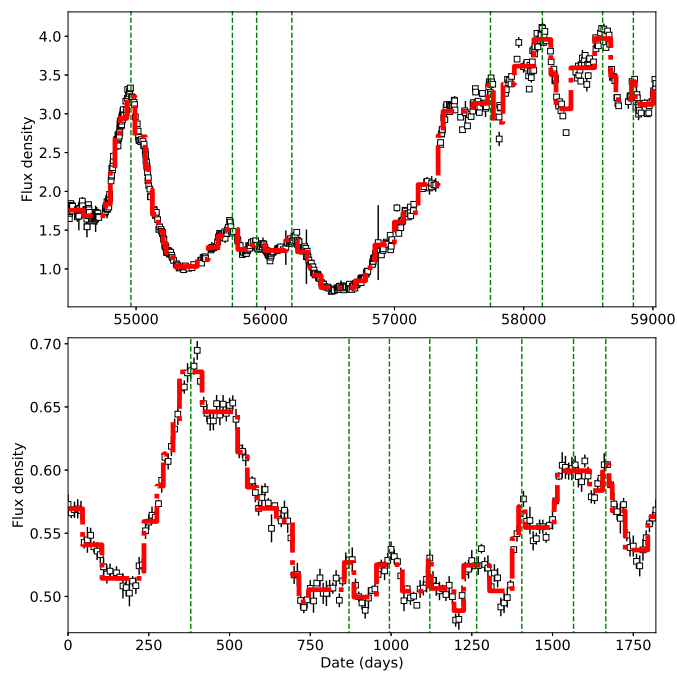}}
\caption{Bayesian block representation (red dash-dotted line) of an observed (CGRaBS~J1504+1029, top panel) and a simulated source (bottom panel). The vertical dashed green lines indicate the location of the identified flares.}
\label{plt:bayesian}
\end{figure}

The above calculations clearly demonstrate how a positional association is not enough to prove the physical connection between jetted-AGN and neutrinos at a high significance. In order to do that, one needs to somehow informatively reduce the number of trials $n$, which can only be achieved by accounting for timing information. In H21 we showed using the CGRaBS sample that the probability for observing a strong flare from a jetted-AGN coincident in time and position with a neutrino event was $p=0.03$ after accounting for all trials. This $2\sigma$ result is by no means conclusive, but certainly an improvement. Although increasing the sample size will increase the probability of chance association, it will also allow one to find more real matches, if they exist, which will improve statistics. Similar conclusions for the need of coincident flares were reached in \cite{Capel2022} looking at the neutrino -- $\gamma$-ray connection through population modeling.

It is important to assess the detectability of a potential correlation with jetted-AGN given the aforementioned phenomenological challenges we face. To gain further insight in the necessary conditions and evaluate the current methodology to achieve our objectives, we simulated neutrino counterpart populations under the null hypothesis that those neutrinos are emitted during radio flares. We start by asking the simple question: can we detect a correlation if all neutrinos are astrophysical and if all were produced during the peaks of radio flares? To address this question we used the radio light curves and the neutrino sample from H21. We used the events in the H21 neutrino sample and assigned new arrival times so that neutrinos coincide with peaks of radio flares. We then repeated the same analysis as in H21. This is achieved by modeling  the light curves of the associated CGRaBS sources found in H21 with Bayesian blocks \citep{Scargle2013}. We identified flares as the local maxima in the Bayesian block representation (green dash lines, Fig. \ref{plt:bayesian}) following \cite{Liodakis2018} and \cite{Liodakis2019}. We used the width of the peak block to randomly assign an arrival time to the neutrino associated with a given source and created 1000 simulated samples. We considered three possibilities. (1) we randomly select identified flares in the light curve; (2) we randomly select flares whose peak blocks have a flux density higher than the median of the flares in that light curve; and (3) we select only the highest flux-density flare. For each of the cases, we repeated the analysis as in H21 and compared our results with the observed. The results for the different simulation cases are given in Table~\ref{table:pvals}. Here we provide a short description and refer the reader to H21 for the details of the cross-correlation analysis. We first start by multiplying the reported neutrino localization RA and DEC uncertainties by 1.3 to convert them to the 90\% sky containment ellipse, and then added the systematic uncertainty in quadrature. Once we have the final localization uncertainty, we cross-matched it with the coordinates of the sources in our sample to find the ones positionally associated with the neutrino region. We estimated the activity index of each ``associated'' source as the ratio of the mean flux-density within a 2-year interval centered around the neutrino arrival time over the mean flux-density of the entire light curve (excluding the neutrino event). We then proceeded to generate simulated samples by randomly shifting the neutrino positions only in RA to reproduce IceCube's sensitivity declination dependence. We cross-matched the ``new'' neutrino positions with the observed sample and calculate the activity index for each newly associated source. We repeated this process $10^4$ times to evaluate the pre-trial chance probability of obtaining an equal or larger value than in the observed population, under the null hypothesis of no association between radio AGN and high-energy neutrinos. Finally, we estimated the post-trial (look-elsewhere effect corrected) probability by treating each simulated population as the observed and repeating the cross-matching analysis described above.

The values in Cols 4-9 of Table~\ref{table:pvals} then give the fraction of those 1000 simulations that give a post-trial p-value smaller than $3\sigma$ or $2\sigma$ for the different tests that were considered in H21. This will allow us to deduce the likelihood of obtaining a significant result under different hypotheses and conditions. We include both $3\sigma$ and $2\sigma$ limits to see when we would expect a result of higher significance than in H21 for the CGRaBS sample ($3\sigma$) and in which cases the significance is similar to what was obtained ($2.3\sigma$ before accounting for the sample trial, $1.6\sigma$ if the number of samples is accounted for). 

The second test considered is the number of flaring sources in the associated sample (Cols 6 and 7). In H21, we defined a source to be in a flaring state if its activity index is larger than 1.29. In this test, we calculate how many such sources there are in the associated sample and compare this to a random population generated as explained above. The significance in H21 for the CGRaBS sample was $2.8\sigma$ before accounting for the sample trial and  $2.2\sigma$ after accounting for the additional sample trial factor.

The final test (Cols 8 and 9) calculates the mean flux density of the associated sample and compares it to the rest of the population. In H21 the significance of this test for the CGRaBS sample was $2.0\sigma$ before accounting for the sample trial and  $1.2\sigma$ after accounting for the additional sample trial factor.

For the three simulated samples described above we find that if the neutrino arrives during any randomly selected flares without any flux density cutoffs (line 1 of Table~\ref{table:pvals}), only a relatively small fraction of simulations would give us a $2\sigma$ result and especially in the number of flaring sources and mean flux density we would almost never obtain a $3\sigma$ result. Given that our real CGRaBS data in H21 gave us a $\sim3\sigma$ result for all these tests (without considering the additional sample trials), this scenario is physically unlikely. On the other hand, if all the neutrinos were astrophysical and arrived during the peak of the highest flare in the source (line 3 of Table~\ref{table:pvals}) we should have obtained a $3\sigma$ result also in H21, which was not the case so from these three tests considered, the most realistic one is the case where the neutrino arrives during a flare with an amplitude higher than the median amplitude (line 2), as this gives a relatively high fraction of $2\sigma$ results, similar to the real data in H21.

We then relaxed the assumption that all neutrinos are astrophysical. To achieve that we used the reported signalness of the events \citep{signalness2021}\footnote{\url{https://gcn.gsfc.nasa.gov/amon_icecube_gold_bronze_events.html}}$^,$\footnote{ \url{https://gcn.gsfc.nasa.gov/amon_ehe_events.html}}$^,$\footnote{ \url{https://gcn.gsfc.nasa.gov/amon_hese_events.html}}. The signalness is a measure of the probability that the event is of astrophysical origins. It is defined as
\begin{equation}
Signalness= \frac{N_S(E,\delta)}{N_S(E,\delta) + N_B(E,\delta)},
\end{equation}
where $N_B$ and $N_S$ are the number of background and signal events at a given declination ($\delta$) higher than a proxy-energy $E$  \citep{Blaufuss2019}.
In case a signalness value is not reported, we assumed a 50\% probability of astrophysical origins. We followed the same procedure as above with the exception that when a neutrino is associated we draw a random value from a uniform distribution between [0,1]. If that value is lower than the fractional signalness we consider that neutrino to be ``astrophysical'' and assign it to a flare as described above. If the neutrino is deemed ``atmospheric'' we assigned a random arrival time within the observed light curve. Including the signalness (lines 4 and 5 of Table~\ref{table:pvals}), we find that the median flare case gives a $2\sigma$ result in less than half of the cases, especially in the number of flaring sources test (Col 7), which gave the highest significance in H21. Even in the case of the neutrino arriving during the highest flare, the fraction of $3\sigma$ results is quite small. This indicates that our sample of 1158 was insufficient to obtain a significant result in the first place.

Next we explored if a sample of 5000 sources monitored for 5 years will be sufficient to produce a correlation. The choice for the number of sources and monitoring period is again motivated by the on-going OVRO experiment\footnote{\url{https://sites.astro.caltech.edu/ovroblazars/}}, which we used to build our simulated samples. We used the Timmer \& Koenig method \citep{Timmer1995} to produce simulated light curves following \cite{Nilsson2018}. The model assumes a Gaussian probability density function of the time series and does not include flattening of the spectrum at low frequencies. We also did not include a white noise component in our model. The power spectral density (PSD) slope ($\beta$) values necessary to produce the simulated light curves were randomly drawn from the empirical cumulative distribution function taken from  \cite{maxmoerbeck2014}. 

About 24\% of the sources in the real OVRO sample are faint ($<$60mJy) and near the limit where variability can be detected in the OVRO light curves \citep[][]{Richards2011}. The light curves of those sources typically exhibit near white noise type variability. We assumed the PSDs of those sources follow an exponential distribution with a mean of 0.175. This results in a PSD distribution bounded between [0,1.65]. We generate $\sim 40$ year long light curves from which we randomly select a 5-year interval. We used the observed flux densities and uncertainties for randomly selected sources to rescale our simulated light curves to the observed range. The light curves were then sampled to the 10-day cadence of the current OVRO monitoring program and the measurement uncertainties were added. An example of the final simulated light curves is shown in Fig. \ref{plt:bayesian} (bottom panel). 

Given the current detection rate of all ``gold'' neutrino events (i.e., events with $>$50\% probability of astrophysical origins), we expect roughly 60 events over five years \citep{Blaufuss2019}. Therefore we used the 56 neutrinos listed in H21 along with their sky positions, localization uncertainties, and signalness. We randomly assigned coordinates from the observed OVRO sample to the simulated light curves and then associated the 5000 source sample with the neutrino positions. This results in 32 neutrinos associated with the simulated sample. In 22 cases there are multiple sources within the localization uncertainty of individual neutrinos. We then repeated the analysis discussed above considering only the cases where the signalness is accounted for in the simulated samples. The results of this test are given on lines 6  and 7 of Table~\ref{table:pvals}. In these cases, the fraction of $2\sigma$ results is fairly high especially when considering neutrinos arriving during the maximum flares, but it would be very unlikely for us to ever obtain a $3\sigma$ result and improve the statistics of H21 because it would also require (unrealistically) that the neutrinos arrive only during the highest flares.

So far we have assumed that neutrino emission is independent of radio flux density. To investigate a possible dependence,  we replaced the ``associated'' sources in our simulated sample with randomly selected sources that have a median flux density of $>0.5$~Jy and $>1$~Jy. These tests are given on lines 8-11 of Table~\ref{table:pvals}. These results show that if the neutrinos arrived during the largest flares, we would very likely get a result that is significant at the $3\sigma$ level, especially if it is the brightest radio sources that are responsible for the neutrino emission, as claimed by \cite{2020ApJ...894..101P}. Even the case where the neutrino arrives during brighter than median flares would give a significant result in 40\% of the simulations when looking at the number of flaring sources, demonstrating that with a large enough sample it will be possible to obtain more conclusive results.

\section{Conclusions}\label{sec:concl}

In this paper we discussed strategies that can potentially allow us to discover whether jetted-AGN are neutrino emitters. We focused on the case where all high-energy $\sim$PeV neutrinos are originating from jetted-AGN. In summary, monitoring of a large, statistically well defined, unbiased sample is necessary if we are to have any chance of proving or disproving a possible correlation.  To further test the capabilities of our current methodology we compared the results of the observed correlation with simulated neutrino samples. We first tested a scenario with existing data used in H21 and artificially placing neutrinos during radio flares. We found that if all neutrinos are of astrophysical origin and arrive at the peak of the brightest radio flares we would have detected a significant correlation. However, we know for a fact that neutrinos do not always arrive at the peak of the flares. Placing neutrinos at any radio flare produces less significant results than in the observed case. Including the signalness in our simulations we find that even in the unreasonable case of all neutrinos coming at the peak of the brightest flares we would most likely not detect a significant correlation. This is in part due to the sample size and the number of detected/associated neutrinos, but also in part due to the uncertainty of the neutrinos' origin. For seven out of the associated neutrinos in H21 we have assumed a 50\% signalness, and three have a reported signalness of $<50\%$. Given that those estimates come from detector limitations, location of the conversion point etc. it is likely that they do not represent the true probability of astrophysical origin. Therefore, it is likely we are polluting our simulated samples with false negatives. 

We then explored if a larger sample can produce significant results. We simulated 5,000 -- 5-year long -- light curves which were then associated with the existing high-energy neutrino sample. We then repeated the above analysis accounting for the signalness of the neutrinos. Assuming neutrinos arrive at the peak of the brightest flares, regardless of radio brightness, then our results would not be significantly improved over H21. On the other hand, if only the brightest radio sources are neutrino emitters, then the current OVRO experiment will most likely find a significant correlation. This is particularly important given the clear relation between radio brightness and synchrotron peak (HSPs are typically $<0.5$Jy). Currently, no more than one high-energy neutrino has been found within the same localization uncertainty. Detecting multiple neutrinos coming from the same sources would tremendously improve the statistical significance of our analysis. This would most likely require a longer period than the one considered for our simulations, highlighting the need for continuous long-term monitoring programs. As previously discussed, a larger sample would increase the chance coincident, but also the real associations. Unfortunately, samples $>5000$ sources are beyond the monitoring capabilities of any existing single radio telescope with a reasonable cadence. Upcoming large scale radio surveys (e.g., Simons observatory -- \citealp{Ade2019_simons}, CMB-S4 -- \citealp{CMBS42022}) can provide the necessary monitoring capabilities, and would require neutrino observatories in the northern hemisphere. While our analysis is focused on the radio regime, the overall strategies are applicable to other wavelengths. Vera C. Rubin observatory's legacy survey of space and time (LSST) will be soon monitoring millions of AGN \citep{LSST2009}. Assuming a linear relation between neutrinos and bolometric luminosity, \cite{Creque-Sarbinowski2022} found that a statistically significant correlation can be achieved for 2.8$\times10^7$ AGN monitored for the nominal 10-year duration of LSST, even if only a small fraction of AGN produce neutrinos. The authors find that the correlation significantly improves when considering AGN variability and using neutrino detectors in the opposite hemisphere, well inline with the strategies outlined here.

We note that in our analysis considering the signalness we assumed that all astrophysical neutrinos originate in jetted-AGN. If different types of sources are responsible for the high-energy neutrinos observed by IceCube, in addition to AGN (e.g., TDEs,  \citealp{vanvelzen2021}), it will have a negative impact on the significance of the correlation. From our analysis we can come to the following conclusions.
\begin{itemize}

\item Our simulations of neutrinos arriving at any radio flare (possibility (1) in section \ref{sec:sims}) yielded a less significant correlation than that which was observed. This suggests that there is a positive relation between neutrinos and radio brightness.

\item Similarly, our results imply that at least some of the neutrinos that appear to be associated with blazars are likely astrophysical.

\item The sample in H21, although the largest timing study thus far, was not sufficient to yield a significant correlation. Larger samples are necessary.

\item A $\sim$5,000 source sample monitored over 5 years is sufficient to detect a correlation at a $>3\sigma$ level, only if the brightest ($>$0.5~Jy) radio sources emit neutrinos.

\item If all blazars are equally good neutrino emitters, it is unlikely a correlation more statistically significant than then one reported in H21 will be found.

\item Although not definitive, a less significant correlation than the one reported in H21 would argue against the blazar-neutrino association.

\end{itemize}

Improving the state-of-the-art analysis tools (e.g., an improved measure for the activity index), as well as new observations from operating (ANTARES, \citealp{Illuminati2021}; Baikal-GVD, \citealp{Belolaptikov2021}) neutrino telescopes will help us mitigate the current observational limitations. Future facilities such as KM3NET \citep{Aiello2019,Km3net2016} as well as upgrades to IceCube \citep[IceCube-gen2,][]{icecube-gen2_2021} will significantly improve the angular resolution and the accuracy of the reconstructed arrival direction. The sub-degree accuracy that could be achieved for high-energy neutrinos ($>10$~TeV) will be instrumental in reducing the number of jetted-AGN within the localization uncertainty, and therefore improving the significance of spatial and temporal correlations, as well as revealing neutrino point-sources by associating multiple neutrinos to a single jetted-AGN.

\begin{acknowledgements}
The authors thank Keith Grainge and the anonymous referee for comments and suggestions that helped improve this work. T. H. was supported by the Academy of Finland projects 317383, 320085, 322535, and 345899.
This research has made use of data from the OVRO 40-m monitoring program \citep{Richards2011}, supported by private funding from the California Institute of Technology and the Max Planck Institute for Radio Astronomy, and by NASA grants NNX08AW31G, NNX11A043G, and NNX14AQ89G and NSF grants AST-0808050 and AST- 1109911. E. L. was supported by Academy of Finland projects 317636 and 320045. The computer resources of the Finnish IT Center for Science (CSC) and the FGCI project (Finland) are acknowledged. S.K. acknowledges support from the European Research Council (ERC) under the European Unions Horizon 2020 research and innovation programme under grant agreement No.~771282. W.M. gratefully acknowledges support by the ANID BASAL projects ACE210002 and FB210003, and FONDECYT 11190853. V.P. acknowledges support by the Hellenic Foundation for Research and Innovation (H.F.R.I.) under the “First Call for H.F.R.I. Research Projects to support Faculty members and Researchers and the procurement of high-cost research equipment grant” (Project 1552 CIRCE), and from the Foundation of Research and Technology - Hellas Synergy Grants Program through project MagMASim, jointly implemented by the Institute of Astrophysics and the Institute of Applied and Computational Mathematics.
\end{acknowledgements}

%

\begin{thebibliography}{68}
\expandafter\ifx\csname natexlab\endcsname\relax\def\natexlab#1{#1}\fi

\bibitem[{{Aartsen} {et~al.}(2021){Aartsen}, {Abbasi}, {Ackermann}, {Adams},
  {Aguilar}, {Ahlers}, {Ahrens}, {Alispach}, {Allison}, {Amin}, {Andeen},
  {Anderson}, {Ansseau}, {Anton}, {Arg{\"u}elles}, {Arlen}, {Auffenberg},
  {Axani}, {Bagherpour}, {Bai}, {Balagopal V}, {Barbano}, {Bartos}, {Bastian},
  {Basu}, {Baum}, {Baur}, {Bay}, {Beatty}, {Becker}, {Tjus}, {BenZvi},
  {Berley}, {Bernardini}, {Besson}, {Binder}, {Bindig}, {Blaufuss}, {Blot},
  {Bohm}, {Bohmer}, {B{\"o}ser}, {Botner}, {B{\"o}ttcher}, {Bourbeau},
  {Bourbeau}, {Bradascio}, {Braun}, {Bron}, {Brostean-Kaiser}, {Burgman},
  {Burley}, {Buscher}, {Busse}, {Bustamante}, {Campana}, {Carnie-Bronca},
  {Carver}, {Chen}, {Chen}, {Cheung}, {Chirkin}, {Choi}, {Clark}, {Clark},
  {Classen}, {Coleman}, {Collin}, {Connolly}, {Conrad}, {Coppin}, {Correa},
  {Cowen}, {Cross}, {Dave}, {Deaconu}, {De Clercq}, {DeLaunay}, {De Kockere},
  {Dembinski}, {Deoskar}, {De Ridder}, {Desai}, {Desiati}, {de Vries}, {de
  Wasseige}, {de With}, {DeYoung}, {Dharani}, {Diaz}, {D{\'\i}az-V{\'e}lez},
  {Dujmovic}, {Dunkman}, {DuVernois}, {Dvorak}, {Ehrhardt}, {Eller}, {Engel},
  {Evans}, {Evenson}, {Fahey}, {Farrag}, {Fazely}, {Felde}, {Fienberg},
  {Filimonov}, {Finley}, {Fischer}, {Fox}, {Franckowiak}, {Friedman}, {Fritz},
  {Gaisser}, {Gallagher}, {Ganster}, {Garcia-Fernandez}, {Garrappa}, {Gartner},
  {Gerhard}, {Gernhaeuser}, {Ghadimi}, {Glaser}, {Glauch}, {Gl{\"u}senkamp},
  {Goldschmidt}, {Gonzalez}, {Goswami}, {Grant}, {Gr{\'e}goire}, {Griffith},
  {Griswold}, {G{\"u}nd{\"u}z}, {Haack}, {Hallgren}, {Halliday}, {Halve},
  {Halzen}, {Hanson}, {Hanson}, {Hardin}, {Haugen}, {Haungs}, {Hauser},
  {Hebecker}, {Heinen}, {Heix}, {Helbing}, {Hellauer}, {Henningsen},
  {Hickford}, {Hignight}, {Hill}, {Hill}, {Hoffman}, {Hoffmann}, {Hoffmann},
  {Hoinka}, {Hokanson-Fasig}, {Holzapfel}, {Hoshina}, {Huang}, {Huber},
  {Huber}, {Huege}, {Hughes}, {Hultqvist}, {H{\"u}nnefeld}, {Hussain}, {In},
  {Iovine}, {Ishihara}, {Jansson}, {Japaridze}, {Jeong}, {Jones}, {Jonske},
  {Joppe}, {Kalekin}, {Kang}, {Kang}, {Kang}, {Kappes}, {Kappesser}, {Karg},
  {Karl}, {Karle}, {Katori}, {Katz}, {Kauer}, {Keivani}, {Kellermann},
  {Kelley}, {Kheirandish}, {Kim}, {Kin}, {Kintscher}, {Kiryluk}, {Kittler},
  {Kleifges}, {Klein}, {Koirala}, {Kolanoski}, {K{\"o}pke}, {Kopper}, {Kopper},
  {Koskinen}, {Koundal}, {Kovacevich}, {Kowalski}, {Krauss}, {Krings},
  {Kr{\"u}ckl}, {Kulacz}, {Kurahashi}, {Gualda}, {Lahmann}, {Lanfranchi},
  {Larson}, {Latif}, {Lauber}, {Lazar}, {Leonard}, {Leszczy{\'n}ska}, {Li},
  {Liu}, {Lohfink}, {LoSecco}, {Mariscal}, {Lu}, {Lucarelli}, {Ludwig},
  {L{\"u}nemann}, {Luszczak}, {Lyu}, {Ma}, {Madsen}, {Maggi}, {Mahn}, {Makino},
  {Mallik}, {Mancina}, {Mandalia}, {Mari{\c{s}}}, {Marka}, {Marka}, {Maruyama},
  {Mase}, {Maunu}, {McNally}, {Meagher}, {Medina}, {Meier}, {Meighen-Berger},
  {Merz}, {Meyers}, {Micallef}, {Mockler}, {Moment{\'e}}, {Montaruli}, {Moore},
  {Morse}, {Moulai}, {Muth}, {Naab}, {Nagai}, {Nam}, {Nauman}, {Necker},
  {Neer}, {Nelles}, {Nguyễn}, {Niederhausen}, {Nisa}, {Nowicki}, {Nygren},
  {Oberla}, {Pollmann}, {Oehler}, {Olivas}, {O'Sullivan}, {Pan}, {Pandya},
  {Pankova}, {Papp}, {Park}, {Parker}, {Paudel}, {Peiffer}, {P{\'e}rez de los
  Heros}, {Petersen}, {Philippen}, {Pieloth}, {Pieper}, {Pinfold}, {Pizzuto},
  {Plaisier}, {Plum}, {Popovych}, {Porcelli}, {Rodriguez}, {Price},
  {Przybylski}, {Raab}, {Raissi}, {Rameez}, {Rauch}, {Rawlins}, {Rea},
  {Rehman}, {Reimann}, {Renschler}, {Renzi}, {Resconi}, {Reusch}, {Rhode},
  {Richman}, {Riedel}, {Riegel}, {Roberts}, {Robertson}, {Roellinghoff},
  {Rongen}, {Rott}, {Ruhe}, {Ryckbosch}, {Cantu}, {Safa}, {Herrera},
  {Sandrock}, {Sandroos}, {Sandstrom}, {Santander}, {Sarkar}, {Sarkar},
  {Satalecka}, {Scharf}, {Schaufel}, {Schieler}, {Schlunder}, {Schmidt},
  {Schneider}, {Schneider}, {Schr{\"o}der}, {Schumacher}, {Sclafani}, {Seckel},
  {Seunarine}, {Shaevitz}, {Sharma}, {Shefali}, {Silva}, {Smith}, {Smithers},
  {Snihur}, {Soedingrekso}, {Soldin}, {S{\"o}ldner-Rembold}, {Song},
  {Southall}, {Spiczak}, {Spiering}, {Stachurska}, {Stamatikos}, {Stanev},
  {Stein}, {Stettner}, {Steuer}, {Stezelberger}, {Stokstad}, {Strotjohann},
  {St{\"u}rwald}, {Stuttard}, {Sullivan}, {Taboada}, {Taketa}, {Tanaka},
  {Tenholt}, {Ter-Antonyan}, {Terliuk}, {Tilav}, {Tollefson}, {Tomankova},
  {T{\"o}nnis}, {Torres}, {Toscano}, {Tosi}, {Trettin}, {Tselengidou}, {Tung},
  {Turcati}, {Turcotte}, {Turley}, {Twagirayezu}, {Ty}, {Unger}, {Elorrieta},
  {Vandenbroucke}, {van Eijk}, {van Eijndhoven}, {Vannerom}, {van Santen},
  {Veberic}, {Verpoest}, {Vieregg}, {Vraeghe}, {Walck}, {Watson}, {Weaver},
  {Weindl}, {Weinstock}, {Weiss}, {Weldert}, {Welling}, {Wendt}, {Werthebach},
  {Whitehorn}, {Wiebe}, {Wiebusch}, {Williams}, {Wissel}, {Wolf}, {Wood},
  {Woschnagg}, {Wrede}, {Wren}, {Wulff}, {Xu}, {Xu}, {Yanez}, {Yoshida},
  {Yuan}, {Zhang}, {Zierke}, \& {Z{\"o}cklein}}]{icecube-gen2_2021}
{Aartsen}, M.~G., {Abbasi}, R., {Ackermann}, M., {et~al.} 2021, Journal of
  Physics G Nuclear Physics, 48, 060501

\bibitem[{{Aartsen} {et~al.}(2017){Aartsen}, {Abraham}, {Ackermann}, {Adams},
  {Aguilar}, {Ahlers}, {Ahrens}, {Altmann}, {Andeen}, {Anderson}, {Ansseau},
  {Anton}, {Archinger}, {Arguelles}, {Arlen}, {Auffenberg}, {Axani}, {Bai},
  {Barwick}, {Baum}, {Bay}, {Beatty}, {Becker Tjus}, {Becker}, {BenZvi},
  {Berghaus}, {Berley}, {Bernardini}, {Bernhard}, {Besson}, {Binder}, {Bindig},
  {Bissok}, {Blaufuss}, {Blot}, {Boersma}, {Bohm}, {B{\"o}rner}, {Bos}, {Bose},
  {B{\"o}ser}, {Botner}, {Braun}, {Brayeur}, {Bretz}, {Burgman}, {Casey},
  {Casier}, {Cheung}, {Chirkin}, {Christov}, {Clark}, {Classen}, {Coenders},
  {Collin}, {Conrad}, {Cowen}, {Cruz Silva}, {Daughhetee}, {Davis}, {Day}, {de
  Andr{\'e}}, {De Clercq}, {del Pino Rosendo}, {Dembinski}, {De Ridder},
  {Desiati}, {de Vries}, {de Wasseige}, {de With}, {DeYoung},
  {D{\'\i}az-V{\'e}lez}, {di Lorenzo}, {Dujmovic}, {Dumm}, {Dunkman},
  {Eberhardt}, {Ehrhardt}, {Eichmann}, {Euler}, {Evenson}, {Fahey}, {Fazely},
  {Feintzeig}, {Felde}, {Filimonov}, {Finley}, {Flis}, {F{\"o}sig},
  {Franckowiak}, {Fuchs}, {Gaisser}, {Gaior}, {Gallagher}, {Gerhardt},
  {Ghorbani}, {Giang}, {Gladstone}, {Glagla}, {Gl{\"u}senkamp}, {Goldschmidt},
  {Golup}, {Gonzalez}, {G{\'o}ra}, {Grant}, {Griffith}, {Haack}, {Haj Ismail},
  {Hallgren}, {Halzen}, {Hansen}, {Hansmann}, {Hansmann}, {Hanson}, {Hebecker},
  {Heereman}, {Helbing}, {Hellauer}, {Hickford}, {Hignight}, {Hill}, {Hoffman},
  {Hoffmann}, {Holzapfel}, {Homeier}, {Hoshina}, {Huang}, {Huber}, {Huelsnitz},
  {Hultqvist}, {In}, {Ishihara}, {Jacobi}, {Japaridze}, {Jeong}, {Jero},
  {Jones}, {Jurkovic}, {Kappes}, {Karg}, {Karle}, {Katz}, {Kauer}, {Keivani},
  {Kelley}, {Kemp}, {Kheirandish}, {Kim}, {Kintscher}, {Kiryluk}, {Kittler},
  {Klein}, {Kohnen}, {Koirala}, {Kolanoski}, {Konietz}, {K{\"o}pke}, {Kopper},
  {Kopper}, {Koskinen}, {Kowalski}, {Krings}, {Kroll}, {Kr{\"u}ckl},
  {Kr{\"u}ger}, {Kunnen}, {Kunwar}, {Kurahashi}, {Kuwabara}, {Labare},
  {Lanfranchi}, {Larson}, {Lennarz}, {Lesiak-Bzdak}, {Leuermann}, {Leuner},
  {Lu}, {L{\"u}nemann}, {Madsen}, {Maggi}, {Mahn}, {Mancina}, {Mandelartz},
  {Maruyama}, {Mase}, {Maunu}, {McNally}, {Meagher}, {Medici}, {Meier}, {Meli},
  {Menne}, {Merino}, {Meures}, {Miarecki}, {Middell}, {Mohrmann}, {Montaruli},
  {Moulai}, {Nahnhauer}, {Naumann}, {Neer}, {Niederhausen}, {Nowicki},
  {Nygren}, {Obertacke Pollmann}, {Olivas}, {Omairat}, {O'Murchadha},
  {Palczewski}, {Pandya}, {Pankova}, {Penek}, {Pepper}, {P{\'e}rez de los
  Heros}, {Pfendner}, {Pieloth}, {Pinat}, {Posselt}, {Price}, {Przybylski},
  {Quinnan}, {Raab}, {R{\"a}del}, {Rameez}, {Rawlins}, {Reimann}, {Relich},
  {Resconi}, {Rhode}, {Richman}, {Riedel}, {Robertson}, {Rongen}, {Rott},
  {Ruhe}, {Ryckbosch}, {Rysewyk}, {Sabbatini}, {Sanchez Herrera}, {Sandrock},
  {Sandroos}, {Sarkar}, {Satalecka}, {Schimp}, {Schlunder}, {Schmidt},
  {Schoenen}, {Sch{\"o}neberg}, {Sch{\"o}nwald}, {Schumacher}, {Seckel},
  {Seunarine}, {Soldin}, {Song}, {Spiczak}, {Spiering}, {Stahlberg},
  {Stamatikos}, {Stanev}, {Stasik}, {Steuer}, {Stezelberger}, {Stokstad},
  {St{\"o}{\ss}l}, {Str{\"o}m}, {Strotjohann}, {Sullivan}, {Sutherland},
  {Taavola}, {Taboada}, {Tatar}, {Ter-Antonyan}, {Terliuk}, {Te{\v{s}}i{\'c}},
  {Tilav}, {Toale}, {Tobin}, {Toscano}, {Tosi}, {Tselengidou}, {Turcati},
  {Unger}, {Usner}, {Vallecorsa}, {Vandenbroucke}, {van Eijndhoven},
  {Vanheule}, {van Rossem}, {van Santen}, {Veenkamp}, {Vehring}, {Voge},
  {Vraeghe}, {Walck}, {Wallace}, {Wallraff}, {Wandkowsky}, {Weaver}, {Wendt},
  {Westerhoff}, {Whelan}, {Wickmann}, {Wiebe}, {Wiebusch}, {Wille}, {Williams},
  {Wills}, {Wissing}, {Wolf}, {Wood}, {Woolsey}, {Woschnagg}, {Xu}, {Xu}, {Xu},
  {Yanez}, {Yodh}, {Yoshida}, {Zoll}, \& {IceCube
  Collaboration}}]{2017ApJ...835...45A}
{Aartsen}, M.~G., {Abraham}, K., {Ackermann}, M., {et~al.} 2017, \apj, 835, 45

\bibitem[{{Aartsen} {et~al.}(2014){Aartsen}, {Ackermann}, {Adams}, {Aguilar},
  {Ahlers}, {Ahrens}, {Altmann}, {Anderson}, {Arguelles}, {Arlen},
  {Auffenberg}, {Bai}, {Barwick}, {Baum}, {Beatty}, {Becker Tjus}, {Becker},
  {BenZvi}, {Berghaus}, {Berley}, {Bernardini}, {Bernhard}, {Besson}, {Binder},
  {Bindig}, {Bissok}, {Blaufuss}, {Blumenthal}, {Boersma}, {Bohm}, {Bose},
  {B{\"o}ser}, {Botner}, {Brayeur}, {Bretz}, {Brown}, {Casey}, {Casier},
  {Chirkin}, {Christov}, {Christy}, {Clark}, {Classen}, {Clevermann},
  {Coenders}, {Cowen}, {Cruz Silva}, {Danninger}, {Daughhetee}, {Davis}, {Day},
  {de Andr{\'e}}, {De Clercq}, {De Ridder}, {Desiati}, {de Vries}, {de With},
  {DeYoung}, {D{\'\i}az-V{\'e}lez}, {Dunkman}, {Eagan}, {Eberhardt},
  {Eichmann}, {Eisch}, {Euler}, {Evenson}, {Fadiran}, {Fazely}, {Fedynitch},
  {Feintzeig}, {Felde}, {Feusels}, {Filimonov}, {Finley}, {Fischer-Wasels},
  {Flis}, {Franckowiak}, {Frantzen}, {Fuchs}, {Gaisser}, {Gallagher},
  {Gerhardt}, {Gier}, {Gladstone}, {Gl{\"u}senkamp}, {Goldschmidt}, {Golup},
  {Gonzalez}, {Goodman}, {G{\'o}ra}, {Grandmont}, {Grant}, {Gretskov}, {Groh},
  {Gro{\ss}}, {Ha}, {Haack}, {Haj Ismail}, {Hallen}, {Hallgren}, {Halzen},
  {Hanson}, {Hebecker}, {Heereman}, {Heinen}, {Helbing}, {Hellauer}, {Hellwig},
  {Hickford}, {Hill}, {Hoffman}, {Hoffmann}, {Homeier}, {Hoshina}, {Huang},
  {Huelsnitz}, {Hulth}, {Hultqvist}, {Hussain}, {Ishihara}, {Jacobi},
  {Jacobsen}, {Jagielski}, {Japaridze}, {Jero}, {Jlelati}, {Jurkovic},
  {Kaminsky}, {Kappes}, {Karg}, {Karle}, {Kauer}, {Kelley}, {Kheirandish},
  {Kiryluk}, {Kl{\"a}s}, {Klein}, {K{\"o}hne}, {Kohnen}, {Kolanoski}, {Koob},
  {K{\"o}pke}, {Kopper}, {Kopper}, {Koskinen}, {Kowalski}, {Kriesten},
  {Krings}, {Kroll}, {Kunnen}, {Kurahashi}, {Kuwabara}, {Labare}, {Larsen},
  {Larson}, {Lesiak-Bzdak}, {Leuermann}, {Leute}, {L{\"u}nemann},
  {Mac{\'\i}as}, {Madsen}, {Maggi}, {Maruyama}, {Mase}, {Matis}, {McNally},
  {Meagher}, {Meli}, {Meures}, {Miarecki}, {Middell}, {Middlemas}, {Milke},
  {Miller}, {Mohrmann}, {Montaruli}, {Morse}, {Nahnhauer}, {Naumann},
  {Niederhausen}, {Nowicki}, {Nygren}, {Obertacke}, {Odrowski}, {Olivas},
  {Omairat}, {O'Murchadha}, {Palczewski}, {Paul}, {Penek}, {Pepper}, {P{\'e}rez
  de los Heros}, {Pfendner}, {Pieloth}, {Pinat}, {Posselt}, {Price},
  {Przybylski}, {P{\"u}tz}, {Quinnan}, {R{\"a}del}, {Rameez}, {Rawlins},
  {Redl}, {Rees}, {Reimann}, {Resconi}, {Rhode}, {Richman}, {Riedel},
  {Robertson}, {Rodrigues}, {Rongen}, {Rott}, {Ruhe}, {Ruzybayev}, {Ryckbosch},
  {Saba}, {Sand er}, {Santander}, {Sarkar}, {Schatto}, {Scheriau}, {Schmidt},
  {Schmitz}, {Schoenen}, {Sch{\"o}neberg}, {Sch{\"o}nwald}, {Schukraft},
  {Schulte}, {Schulz}, {Seckel}, {Sestayo}, {Seunarine}, {Shanidze},
  {Sheremata}, {Smith}, {Soldin}, {Spiczak}, {Spiering}, {Stamatikos},
  {Stanev}, {Stanisha}, {Stasik}, {Stezelberger}, {Stokstad}, {St{\"o}{\ss}l},
  {Strahler}, {Str{\"o}m}, {Strotjohann}, {Sullivan}, {Taavola}, {Taboada},
  {Tamburro}, {Tepe}, {Ter-Antonyan}, {Terliuk}, {Te{\v{s}}i{\'c}}, {Tilav},
  {Toale}, {Tobin}, {Tosi}, {Tselengidou}, {Unger}, {Usner}, {Vallecorsa}, {van
  Eijndhoven}, {Vandenbroucke}, {van Santen}, {Vehring}, {Voge}, {Vraeghe},
  {Walck}, {Wallraff}, {Weaver}, {Wellons}, {Wendt}, {Westerhoff}, {Whelan},
  {Whitehorn}, {Wichary}, {Wiebe}, {Wiebusch}, {Williams}, {Wissing}, {Wolf},
  {Wood}, {Woschnagg}, {Xu}, {Xu}, {Yanez}, {Yodh}, {Yoshida}, {Zarzhitsky},
  {Ziemann}, {Zierke}, {Zoll}, \& {IceCube
  Collaboration}}]{2014PhRvL.113j1101A}
{Aartsen}, M.~G., {Ackermann}, M., {Adams}, J., {et~al.} 2014, \prl, 113,
  101101

\bibitem[{{Abazajian} {et~al.}(2022){Abazajian}, {Abdulghafour}, {Addison},
  {Adshead}, {Ahmed}, {Ajello}, {Akerib}, {Allen}, {Alonso}, {Alvarez}, {Amin},
  {Amiri}, {Anderson}, {Ansarinejad}, {Archipley}, {Arnold}, {Ashby}, {Aung},
  {Baccigalupi}, {Baker}, {Bakshi}, {Bard}, {Barkats}, {Barron}, {Barry},
  {Bartlett}, {Barton}, {Basu Thakur}, {Battaglia}, {Beall}, {Bean}, {Beck},
  {Belkner}, {Benabed}, {Bender}, {Benson}, {Besuner}, {Bethermin}, {Bhimani},
  {Bianchini}, {Biquard}, {Birdwell}, {Bischoff}, {Bleem}, {Bocaz}, {Bock},
  {Bocquet}, {Boddy}, {Bond}, {Borrill}, {Bouchet}, {Brinckmann}, {Brown},
  {Bryan}, {Buza}, {Byrum}, {Calabrese}, {Calafut}, {Caldwell}, {Carlstrom},
  {Carron}, {Cecil}, {Challinor}, {Chan}, {Chang}, {Chapman}, {Charles},
  {Chauvin}, {Cheng}, {Chesmore}, {Cheung}, {Chinone}, {Chluba}, {Cho}, {Choi},
  {Clancy}, {Clark}, {Cooray}, {Coppi}, {Corlett}, {Coulton}, {Crawford},
  {Crites}, {Cukierman}, {Cyr-Racine}, {Dai}, {Daley}, {Dart}, {Daues}, {de
  Haan}, {Deaconu}, {Delabrouille}, {Derylo}, {Devlin}, {Di Valentino},
  {Dierickx}, {Dober}, {Doriese}, {Duff}, {Dutcher}, {Dvorkin}, {D{\"u}nner},
  {Eftekhari}, {Eimer}, {El Bouhargani}, {Elleflot}, {Emerson}, {Errard},
  {Essinger-Hileman}, {Fabbian}, {Fanfani}, {Fasano}, {Feng}, {Ferraro},
  {Filippini}, {Flauger}, {Flaugher}, {Fraisse}, {Frisch}, {Frolov},
  {Galitzki}, {Gallardo}, {Galli}, {Ganga}, {Gerbino}, {Giannakopoulos},
  {Gilchriese}, {Gluscevic}, {Goeckner-Wald}, {Goldfinger}, {Green}, {Grimes},
  {Grin}, {Grohs}, {Gualtieri}, {Guarino}, {Gudmundsson}, {Gullett}, {Guns},
  {Habib}, {Haller}, {Halpern}, {Halverson}, {Hanany}, {Hand}, {Harrington},
  {Hasegawa}, {Hasselfield}, {Hazumi}, {Heitmann}, {Henderson}, {Hensley},
  {Herbst}, {Hervias-Caimapo}, {Hill}, {Hills}, {Hivon}, {Hlozek}, {Ho},
  {Holder}, {Hollister}, {Holzapfel}, {Hood}, {Hotinli}, {Hryciuk}, {Hubmayr},
  {Huffenberger}, {Hui}, {Ib{\'a} nez}, {Ibitoye}, {Ikape}, {Irwin}, {Jacobus},
  {Jeong}, {Johnson}, {Johnstone}, {Jones}, {Joseph}, {Jost}, {Kang}, {Kaplan},
  {Karkare}, {Katayama}, {Keskitalo}, {King}, {Kisner}, {Klein}, {Knox},
  {Koopman}, {Kosowsky}, {Kovac}, {Kovetz}, {Krolewski}, {Kubik}, {Kuhlmann},
  {Kuo}, {Kusaka}, {L{\"a}hteenm{\"a}ki}, {Lau}, {Lawrence}, {Lee}, {Legrand},
  {Leitner}, {Leloup}, {Lewis}, {Li}, {Linder}, {Liodakis}, {Liu}, {Long},
  {Louis}, {Loverde}, {Lowry}, {Lu}, {Lubin}, {Ma}, {Maccarone},
  {Madhavacheril}, {Maldonado}, {Mantz}, {Marques}, {Matsuda}, {Mauskopf},
  {May}, {McCarrick}, {McCracken}, {McMahon}, {Meerburg}, {Melin}, {Menanteau},
  {Meyers}, {Millea}, {Miranda}, {Mitchell}, {Mohr}, {Moncelsi}, {Monzani},
  {Moshed}, {Mroczkowski}, {Mukherjee}, {M{\"u}nchmeyer}, {Nagai},
  {Nagarajappa}, {Nagy}, {Namikawa}, {Nati}, {Natoli}, {Nerval}, {Newburgh},
  {Nguyen}, {Nichols}, {Nicola}, {Niemack}, {Nord}, {Norton}, {Novosad},
  {O'Brient}, {Omori}, {Orlando}, {Osherson}, {Osten}, {Padin}, {Paine},
  {Partridge}, {Patil}, {Petravick}, {Petroff}, {Pierpaoli}, {Pilleux},
  {Pogosian}, {Prabhu}, {Pryke}, {Puglisi}, {Racine}, {Raghunathan}, {Rahlin},
  {Raveri}, {Reese}, {Reichardt}, {Remazeilles}, {Rizzieri}, {Rocha}, {Roe},
  {Rotermund}, {Roy}, {Ruhl}, {Saba}, {Sailer}, {Salatino}, {Saliwanchik},
  {Sapozhnikov}, {Sathyanarayana Rao}, {Saunders}, {Schaan}, {Schillaci},
  {Schmitt}, {Scott}, {Sehgal}, {Shandera}, {Sherwin}, {Shirokoff}, {Shiu},
  {Simon}, {Singari}, {Slosar}, {Spergel}, {St. Germaine}, {Staggs}, {Stark},
  {Starkman}, {Steinbach}, {Stompor}, {Stoughton}, {Suzuki}, {Tajima},
  {Tandoi}, {Teply}, {Thayer}, {Thompson}, {Thorne}, {Timbie}, {Tomasi},
  {Trendafilova}, {Tristram}, {Tucker}, {Tucker}, {Umilt{\`a}}, {van Engelen},
  {van Marrewijk}, {Vavagiakis}, {Verg{\`e}s}, {Vieira}, {Vieregg}, {Wagoner},
  {Wallisch}, {Wang}, {Wang}, {Watson}, {Watts}, {Weaver}, {Wenzl},
  {Westbrook}, {White}, {Whitehorn}, {Wiedlea}, {Williams}, {Wilson}, {Winch},
  {Wollack}, {Kimmy Wu}, {Xu}, {Yefremenko}, {Yu}, {Zegeye}, {Zivick}, \&
  {Zonca}}]{CMBS42022}
{Abazajian}, K., {Abdulghafour}, A., {Addison}, G.~E., {et~al.} 2022, arXiv
  e-prints, arXiv:2203.08024

\bibitem[{{Abbasi} {et~al.}(2022){Abbasi}, {Ackermann}, {Adams}, {Aguilar},
  {Ahlers}, {Ahrens}, {Alameddine}, {Alispach}, {Alves}, {Amin}, {Andeen},
  {Anderson}, {Anton}, {Arg{\"u}elles}, {Ashida}, {Axani}, {Bai}, {Balagopal
  V.}, {Barbano}, {Barwick}, {Bastian}, {Basu}, {Baur}, {Bay}, {Beatty},
  {Becker}, {Tjus}, {Bellenghi}, {BenZvi}, {Berley}, {Bernardini}, {Besson},
  {Binder}, {Bindig}, {Blaufuss}, {Blot}, {Boddenberg}, {Bontempo}, {Borowka},
  {B{\"o}ser}, {Botner}, {B{\"o}ttcher}, {Bourbeau}, {Bradascio}, {Braun},
  {Brinson}, {Bron}, {Brostean-Kaiser}, {Browne}, {Burgman}, {Burley}, {Busse},
  {Campana}, {Carnie-Bronca}, {Chen}, {Chen}, {Chirkin}, {Choi}, {Clark},
  {Clark}, {Classen}, {Coleman}, {Collin}, {Conrad}, {Coppin}, {Correa},
  {Cowen}, {Cross}, {Dappen}, {Dave}, {De Clercq}, {DeLaunay}, {L{\'o}pez},
  {Dembinski}, {Deoskar}, {Desai}, {Desiati}, {de Vries}, {de Wasseige}, {de
  With}, {DeYoung}, {Diaz}, {D{\'\i}az-V{\'e}lez}, {Dittmer}, {Dujmovic},
  {Dunkman}, {DuVernois}, {Dvorak}, {Ehrhardt}, {Eller}, {Engel}, {Erpenbeck},
  {Evans}, {Evenson}, {Fan}, {Fazely}, {Feigl}, {Fiedlschuster}, {Fienberg},
  {Filimonov}, {Finley}, {Fischer}, {Fox}, {Franckowiak}, {Friedman}, {Fritz},
  {F{\"u}rst}, {Gaisser}, {Gallagher}, {Ganster}, {Garcia}, {Garrappa},
  {Gerhardt}, {Ghadimi}, {Glaser}, {Glauch}, {Gl{\"u}senkamp}, {Gonzalez},
  {Goswami}, {Grant}, {Gr{\'e}goire}, {Griswold}, {G{\"u}nther}, {Gutjahr},
  {Haack}, {Hallgren}, {Halliday}, {Halve}, {Halzen}, {Minh}, {Hanson},
  {Hardin}, {Harnisch}, {Haungs}, {Hebecker}, {Helbing}, {Henningsen},
  {Hettinger}, {Hickford}, {Hignight}, {Hill}, {Hill}, {Hoffman}, {Hoffmann},
  {Hokanson-Fasig}, {Hoshina}, {Huang}, {Huber}, {Huber}, {Hultqvist},
  {H{\"u}nnefeld}, {Hussain}, {Hymon}, {In}, {Iovine}, {Ishihara}, {Jansson},
  {Japaridze}, {Jeong}, {Jin}, {Jones}, {Kang}, {Kang}, {Kang}, {Kappes},
  {Kappesser}, {Kardum}, {Karg}, {Karl}, {Karle}, {Katz}, {Kauer},
  {Kellermann}, {Kelley}, {Kheirandish}, {Kin}, {Kintscher}, {Kiryluk},
  {Klein}, {Koirala}, {Kolanoski}, {Kontrimas}, {K{\"o}pke}, {Kopper},
  {Kopper}, {Koskinen}, {Koundal}, {Kovacevich}, {Kowalski}, {Kozynets}, {Kun},
  {Kurahashi}, {Lad}, {Gualda}, {Lanfranchi}, {Larson}, {Lauber}, {Lazar},
  {Lee}, {Leonard}, {Leszczy{\'n}ska}, {Li}, {Lincetto}, {Liu}, {Liubarska},
  {Lohfink}, {Mariscal}, {Lu}, {Lucarelli}, {Ludwig}, {Luszczak}, {Lyu}, {Ma},
  {Madsen}, {Mahn}, {Makino}, {Mancina}, {Mari{\c{s}}}, {Martinez-Soler},
  {Maruyama}, {Mase}, {McElroy}, {McNally}, {Mead}, {Meagher}, {Mechbal},
  {Medina}, {Meier}, {Meighen-Berger}, {Micallef}, {Mockler}, {Montaruli},
  {Moore}, {Morse}, {Moulai}, {Naab}, {Nagai}, {Naumann}, {Necker}, {Nguyễn},
  {Niederhausen}, {Nisa}, {Nowicki}, {Pollmann}, {Oehler}, {Oeyen}, {Olivas},
  {O'Sullivan}, {Pandya}, {Pankova}, {Park}, {Parker}, {Paudel}, {Paul}, {de
  los Heros}, {Peters}, {Peterson}, {Philippen}, {Pieper}, {Pittermann},
  {Pizzuto}, {Plum}, {Popovych}, {Porcelli}, {Rodriguez}, {Price}, {Pries},
  {Przybylski}, {Raab}, {Raissi}, {Rameez}, {Rawlins}, {Rea}, {Rehman},
  {Reichherzer}, {Reimann}, {Renzi}, {Resconi}, {Reusch}, {Rhode}, {Richman},
  {Riedel}, {Roberts}, {Robertson}, {Roellinghoff}, {Rongen}, {Rott}, {Ruhe},
  {Ryckbosch}, {Cantu}, {Safa}, {Saffer}, {Herrera}, {Sandrock}, {Sandroos},
  {Santander}, {Sarkar}, {Sarkar}, {Satalecka}, {Schaufel}, {Schieler},
  {Schindler}, {Schmidt}, {Schneider}, {Schneider}, {Schr{\"o}der},
  {Schumacher}, {Schwefer}, {Sclafani}, {Seckel}, {Seunarine}, {Sharma},
  {Shefali}, {Silva}, {Skrzypek}, {Smithers}, {Snihur}, {Soedingrekso},
  {Soldin}, {Spannfellner}, {Spiczak}, {Spiering}, {Stachurska}, {Stamatikos},
  {Stanev}, {Stein}, {Stettner}, {Steuer}, {Stezelberger}, {St{\"u}rwald},
  {Stuttard}, {Sullivan}, {Taboada}, {Ter-Antonyan}, {Tilav}, {Tischbein},
  {Tollefson}, {T{\"o}nnis}, {Toscano}, {Tosi}, {Trettin}, {Tselengidou},
  {Tung}, {Turcati}, {Turcotte}, {Turley}, {Twagirayezu}, {Ty}, {Elorrieta},
  {Valtonen-Mattila}, {Vandenbroucke}, {van Eijndhoven}, {Vannerom}, {van
  Santen}, {Verpoest}, {Walck}, {Watson}, {Weaver}, {Weigel}, {Weindl},
  {Weiss}, {Weldert}, {Wendt}, {Werthebach}, {Weyrauch}, {Whitehorn},
  {Wiebusch}, {Williams}, {Wolf}, {Woschnagg}, {Wrede}, {Wulff}, {Xu}, {Yanez},
  {Yoshida}, {Yu}, {Yuan}, {Zhang}, {Zhelnin}, \& {IceCube
  Collaboration}}]{signalness2021}
{Abbasi}, R., {Ackermann}, M., {Adams}, J., {et~al.} 2022, \apj, 928, 50

\bibitem[{{Abbasi} {et~al.}(2021){Abbasi}, {Ackermann}, {Adams}, {Aguilar},
  {Ahlers}, {Ahrens}, {Alispach}, {Alves}, {Amin}, {An}, {Andeen}, {Anderson},
  {Anton}, {Arg{\"u}elles}, {Ashida}, {Axani}, {Bai}, {Balagopal}, {Barbano},
  {Barwick}, {Bastian}, {Basu}, {Baur}, {Bay}, {Beatty}, {Becker}, {Becker
  Tjus}, {Bellenghi}, {BenZvi}, {Berley}, {Bernardini}, {Besson}, {Binder},
  {Bindig}, {Blaufuss}, {Blot}, {Boddenberg}, {Bontempo}, {Borowka},
  {B{\"o}ser}, {Botner}, {B{\"o}ttcher}, {Bourbeau}, {Bradascio}, {Braun},
  {Bron}, {Brostean-Kaiser}, {Browne}, {Burgman}, {Burley}, {Busse}, {Campana},
  {Carnie-Bronca}, {Chen}, {Chirkin}, {Choi}, {Clark}, {Clark}, {Classen},
  {Coleman}, {Collin}, {Conrad}, {Coppin}, {Correa}, {Cowen}, {Cross},
  {Dappen}, {Dave}, {De Clercq}, {DeLaunay}, {Dembinski}, {Deoskar}, {De
  Ridder}, {Desai}, {Desiati}, {de Vries}, {de Wasseige}, {de With}, {DeYoung},
  {Dharani}, {Diaz}, {D{\'\i}az-V{\'e}lez}, {Dittmer}, {Dujmovic}, {Dunkman},
  {DuVernois}, {Dvorak}, {Ehrhardt}, {Eller}, {Engel}, {Erpenbeck}, {Evans},
  {Evenson}, {Fan}, {Fazely}, {Fiedlschuster}, {Fienberg}, {Filimonov},
  {Finley}, {Fischer}, {Fox}, {Franckowiak}, {Friedman}, {Fritz}, {F{\"u}rst},
  {Gaisser}, {Gallagher}, {Ganster}, {Garcia}, {Garrappa}, {Gerhardt},
  {Ghadimi}, {Glaser}, {Glauch}, {Gl{\"u}senkamp}, {Goldschmidt}, {Gonzalez},
  {Goswami}, {Grant}, {Gr{\'e}goire}, {Griswold}, {G{\"u}nd{\"u}z},
  {G{\"u}nther}, {Haack}, {Hallgren}, {Halliday}, {Halve}, {Halzen}, {Ha Minh},
  {Hanson}, {Hardin}, {Harnisch}, {Haungs}, {Hauser}, {Hebecker}, {Helbing},
  {Henningsen}, {Hettinger}, {Hickford}, {Hignight}, {Hill}, {Hill}, {Hoffman},
  {Hoffmann}, {Hoinka}, {Hokanson-Fasig}, {Hoshina}, {Huang}, {Huber}, {Huber},
  {Hultqvist}, {H{\"u}nnefeld}, {Hussain}, {In}, {Iovine}, {Ishihara},
  {Jansson}, {Japaridze}, {Jeong}, {Jones}, {Kang}, {Kang}, {Kang}, {Kappes},
  {Kappesser}, {Karg}, {Karl}, {Karle}, {Katz}, {Kauer}, {Kellermann},
  {Kelley}, {Kheirandish}, {Kin}, {Kintscher}, {Kiryluk}, {Klein}, {Koirala},
  {Kolanoski}, {Kontrimas}, {K{\"o}pke}, {Kopper}, {Kopper}, {Koskinen},
  {Koundal}, {Kovacevich}, {Kowalski}, {Kozynets}, {Kun}, {Kurahashi}, {Lad},
  {Lagunas Gualda}, {Lanfranchi}, {Larson}, {Lauber}, {Lazar}, {Lee},
  {Leonard}, {Leszczy{\'n}ska}, {Li}, {Lincetto}, {Liu}, {Liubarska},
  {Lohfink}, {Mariscal}, {Lu}, {Lucarelli}, {Ludwig}, {Luszczak}, {Lyu}, {Ma},
  {Madsen}, {Mahn}, {Makino}, {Mancina}, {Mari{\c{s}}}, {Maruyama}, {Mase},
  {McElroy}, {McNally}, {Mead}, {Meagher}, {Medina}, {Meier}, {Meighen-Berger},
  {Micallef}, {Mockler}, {Montaruli}, {Moore}, {Morse}, {Moulai}, {Naab},
  {Nagai}, {Naumann}, {Necker}, {Nguy{\^e}n}, {Niederhausen}, {Nisa},
  {Nowicki}, {Nygren}, {Obertacke Pollmann}, {Oehler}, {Oeyen}, {Olivas},
  {O'Sullivan}, {Pandya}, {Pankova}, {Park}, {Parker}, {Paudel}, {Paul},
  {P{\'e}rez de los Heros}, {Peters}, {Peterson}, {Philippen}, {Pieloth},
  {Pieper}, {Pittermann}, {Pizzuto}, {Plum}, {Popovych}, {Porcelli}, {Prado
  Rodriguez}, {Price}, {Pries}, {Przybylski}, {Raab}, {Raissi}, {Rameez},
  {Rawlins}, {Rea}, {Rehman}, {Reichherzer}, {Reimann}, {Renzi}, {Resconi},
  {Reusch}, {Rhode}, {Richman}, {Riedel}, {Roberts}, {Robertson},
  {Roellinghoff}, {Rongen}, {Rott}, {Ruhe}, {Ryckbosch}, {Rysewyk Cantu},
  {Safa}, {Saffer}, {Herrera}, {Sandrock}, {Sandroos}, {Santander}, {Sarkar},
  {Sarkar}, {Satalecka}, {Scharf}, {Schaufel}, {Schieler}, {Schindler},
  {Schlunder}, {Schmidt}, {Schneider}, {Schneider}, {Schr{\"o}der},
  {Schumacher}, {Schwefer}, {Sclafani}, {Seckel}, {Seunarine}, {Sharma},
  {Shefali}, {Silva}, {Skrzypek}, {Smithers}, {Snihur}, {Soedingrekso},
  {Soldin}, {Spannfellner}, {Spiczak}, {Spiering}, {Stachurska}, {Stamatikos},
  {Stanev}, {Stein}, {Stettner}, {Steuer}, {Stezelberger}, {St{\"u}rwald},
  {Stuttard}, {Sullivan}, {Taboada}, {Tenholt}, {Ter-Antonyan}, {Tilav},
  {Tischbein}, {Tollefson}, {Tomankova}, {T{\"o}nnis}, {Toscano}, {Tosi},
  {Trettin}, {Tselengidou}, {Tung}, {Turcati}, {Turcotte}, {Turley},
  {Twagirayezu}, {Ty}, {Unland Elorrieta}, {Valtonen-Mattila}, {Vandenbroucke},
  {van Eijndhoven}, {Vannerom}, {van Santen}, {Verpoest}, {Vraeghe}, {Walck},
  {Watson}, {Weaver}, {Weigel}, {Weindl}, {Weiss}, {Weldert}, {Wendt},
  {Werthebach}, {Weyrauch}, {Whitehorn}, {Wiebusch}, {Williams}, {Wolf},
  {Woschnagg}, {Wrede}, {Wulff}, {Xu}, {Xu}, {Yanez}, {Yoshida}, {Yu}, {Yuan},
  {Zhang}, \& {IceCube Collaboration}}]{Icecube2021}
{Abbasi}, R., {Ackermann}, M., {Adams}, J., {et~al.} 2021, \apjl, 920, L45

\bibitem[{{Abdo} {et~al.}(2010){Abdo}, {Ackermann}, {Agudo}, {Ajello}, {Aller},
  {Aller}, {Angelakis}, {Arkharov}, {Axelsson}, {Bach}, {Baldini}, {Ballet},
  {Barbiellini}, {Bastieri}, {Baughman}, {Bechtol}, {Bellazzini}, {Benitez},
  {Berdyugin}, {Berenji}, {Blandford}, {Bloom}, {Boettcher}, {Bonamente},
  {Borgland}, {Bregeon}, {Brez}, {Brigida}, {Bruel}, {Burnett}, {Burrows},
  {Buson}, {Caliandro}, {Calzoletti}, {Cameron}, {Capalbi}, {Caraveo},
  {Carosati}, {Casandjian}, {Cavazzuti}, {Cecchi}, {{\c{C}}elik}, {Charles},
  {Chaty}, {Chekhtman}, {Chen}, {Chiang}, {Chincarini}, {Ciprini}, {Claus},
  {Cohen-Tanugi}, {Colafrancesco}, {Cominsky}, {Conrad}, {Costamante},
  {Cutini}, {D'ammando}, {Deitrick}, {D'Elia}, {Dermer}, {de Angelis}, {de
  Palma}, {Digel}, {Donnarumma}, {Silva}, {Drell}, {Dubois}, {Dultzin},
  {Dumora}, {Falcone}, {Farnier}, {Favuzzi}, {Fegan}, {Focke}, {Forn{\'e}},
  {Fortin}, {Frailis}, {Fuhrmann}, {Fukazawa}, {Funk}, {Fusco}, {G{\'o}mez},
  {Gargano}, {Gasparrini}, {Gehrels}, {Germani}, {Giebels}, {Giglietto},
  {Giommi}, {Giordano}, {Giuliani}, {Glanzman}, {Godfrey}, {Grenier},
  {Gronwall}, {Grove}, {Guillemot}, {Guiriec}, {Gurwell}, {Hadasch},
  {Hanabata}, {Harding}, {Hayashida}, {Hays}, {Healey}, {Heidt}, {Hiriart},
  {Horan}, {Hoversten}, {Hughes}, {Itoh}, {Jackson}, {J{\'o}hannesson},
  {Johnson}, {Johnson}, {Jorstad}, {Kadler}, {Kamae}, {Katagiri}, {Kataoka},
  {Kawai}, {Kennea}, {Kerr}, {Kimeridze}, {Kn{\"o}dlseder}, {Kocian},
  {Kopatskaya}, {Koptelova}, {Konstantinova}, {Kovalev}, {Kovalev},
  {Kurtanidze}, {Kuss}, {Lande}, {Larionov}, {Latronico}, {Leto}, {Lindfors},
  {Longo}, {Loparco}, {Lott}, {Lovellette}, {Lubrano}, {Madejski}, {Makeev},
  {Marchegiani}, {Marscher}, {Marshall}, {Max-Moerbeck}, {Mazziotta},
  {McConville}, {McEnery}, {Meurer}, {Michelson}, {Mitthumsiri}, {Mizuno},
  {Moiseev}, {Monte}, {Monzani}, {Morselli}, {Moskalenko}, {Murgia},
  {Nestoras}, {Nilsson}, {Nizhelsky}, {Nolan}, {Norris}, {Nuss}, {Ohsugi},
  {Ojha}, {Omodei}, {Orlando}, {Ormes}, {Osborne}, {Ozaki}, {Pacciani},
  {Padovani}, {Pagani}, {Page}, {Paneque}, {Panetta}, {Parent}, {Pasanen},
  {Pavlidou}, {Pelassa}, {Pepe}, {Perri}, {Pesce-Rollins}, {Piranomonte},
  {Piron}, {Pittori}, {Porter}, {Puccetti}, {Rahoui}, {Rain{\`o}}, {Raiteri},
  {Rando}, {Razzano}, {Reimer}, {Reimer}, {Reposeur}, {Richards}, {Ritz},
  {Rochester}, {Rodriguez}, {Romani}, {Ros}, {Roth}, {Roustazadeh}, {Ryde},
  {Sadrozinski}, {Sadun}, {Sanchez}, {Sander}, {Saz Parkinson}, {Scargle},
  {Sellerholm}, {Sgr{\`o}}, {Shaw}, {Sigua}, {Siskind}, {Smith}, {Smith},
  {Spandre}, {Spinelli}, {Starck}, {Stevenson}, {Stratta}, {Strickman},
  {Suson}, {Tajima}, {Takahashi}, {Takahashi}, {Takalo}, {Tanaka}, {Thayer},
  {Thayer}, {Thompson}, {Tibaldo}, {Torres}, {Tosti}, {Tramacere}, {Uchiyama},
  {Usher}, {Vasileiou}, {Verrecchia}, {Vilchez}, {Villata}, {Vitale}, {Waite},
  {Wang}, {Winer}, {Wood}, {Ylinen}, {Zensus}, {Zhekanis}, \&
  {Ziegler}}]{Abdo2010}
{Abdo}, A.~A., {Ackermann}, M., {Agudo}, I., {et~al.} 2010, \apj, 716, 30

\bibitem[{{Ade} {et~al.}(2019){Ade}, {Aguirre}, {Ahmed}, {Aiola}, {Ali},
  {Alonso}, {Alvarez}, {Arnold}, {Ashton}, {Austermann}, {Awan}, {Baccigalupi},
  {Baildon}, {Barron}, {Battaglia}, {Battye}, {Baxter}, {Bazarko}, {Beall},
  {Bean}, {Beck}, {Beckman}, {Beringue}, {Bianchini}, {Boada}, {Boettger},
  {Bond}, {Borrill}, {Brown}, {Bruno}, {Bryan}, {Calabrese}, {Calafut},
  {Calisse}, {Carron}, {Challinor}, {Chesmore}, {Chinone}, {Chluba}, {Cho},
  {Choi}, {Coppi}, {Cothard}, {Coughlin}, {Crichton}, {Crowley}, {Crowley},
  {Cukierman}, {D'Ewart}, {D{\"u}nner}, {de Haan}, {Devlin}, {Dicker},
  {Didier}, {Dobbs}, {Dober}, {Duell}, {Duff}, {Duivenvoorden}, {Dunkley},
  {Dusatko}, {Errard}, {Fabbian}, {Feeney}, {Ferraro}, {Flux{\`a}}, {Freese},
  {Frisch}, {Frolov}, {Fuller}, {Fuzia}, {Galitzki}, {Gallardo}, {Tomas Galvez
  Ghersi}, {Gao}, {Gawiser}, {Gerbino}, {Gluscevic}, {Goeckner-Wald}, {Golec},
  {Gordon}, {Gralla}, {Green}, {Grigorian}, {Groh}, {Groppi}, {Guan},
  {Gudmundsson}, {Han}, {Hargrave}, {Hasegawa}, {Hasselfield}, {Hattori},
  {Haynes}, {Hazumi}, {He}, {Healy}, {Henderson}, {Hervias-Caimapo}, {Hill},
  {Hill}, {Hilton}, {Hilton}, {Hincks}, {Hinshaw}, {Hlo{\v{z}}ek}, {Ho}, {Ho},
  {Howe}, {Huang}, {Hubmayr}, {Huffenberger}, {Hughes}, {Ijjas}, {Ikape},
  {Irwin}, {Jaffe}, {Jain}, {Jeong}, {Kaneko}, {Karpel}, {Katayama}, {Keating},
  {Kernasovskiy}, {Keskitalo}, {Kisner}, {Kiuchi}, {Klein}, {Knowles},
  {Koopman}, {Kosowsky}, {Krachmalnicoff}, {Kuenstner}, {Kuo}, {Kusaka},
  {Lashner}, {Lee}, {Lee}, {Leon}, {Leung}, {Lewis}, {Li}, {Li}, {Limon},
  {Linder}, {Lopez-Caraballo}, {Louis}, {Lowry}, {Lungu}, {Madhavacheril},
  {Mak}, {Maldonado}, {Mani}, {Mates}, {Matsuda}, {Maurin}, {Mauskopf}, {May},
  {McCallum}, {McKenney}, {McMahon}, {Meerburg}, {Meyers}, {Miller},
  {Mirmelstein}, {Moodley}, {Munchmeyer}, {Munson}, {Naess}, {Nati},
  {Navaroli}, {Newburgh}, {Nguyen}, {Niemack}, {Nishino}, {Orlowski-Scherer},
  {Page}, {Partridge}, {Peloton}, {Perrotta}, {Piccirillo}, {Pisano},
  {Poletti}, {Puddu}, {Puglisi}, {Raum}, {Reichardt}, {Remazeilles},
  {Rephaeli}, {Riechers}, {Rojas}, {Roy}, {Sadeh}, {Sakurai}, {Salatino},
  {Sathyanarayana Rao}, {Schaan}, {Schmittfull}, {Sehgal}, {Seibert}, {Seljak},
  {Sherwin}, {Shimon}, {Sierra}, {Sievers}, {Sikhosana}, {Silva-Feaver},
  {Simon}, {Sinclair}, {Siritanasak}, {Smith}, {Smith}, {Spergel}, {Staggs},
  {Stein}, {Stevens}, {Stompor}, {Suzuki}, {Tajima}, {Takakura}, {Teply},
  {Thomas}, {Thorne}, {Thornton}, {Trac}, {Tsai}, {Tucker}, {Ullom},
  {Vagnozzi}, {van Engelen}, {Van Lanen}, {Van Winkle}, {Vavagiakis},
  {Verg{\`e}s}, {Vissers}, {Wagoner}, {Walker}, {Ward}, {Westbrook},
  {Whitehorn}, {Williams}, {Williams}, {Wollack}, {Xu}, {Yu}, {Yu}, {Zago},
  {Zhang}, {Zhu}, \& {Simons Observatory Collaboration}}]{Ade2019_simons}
{Ade}, P., {Aguirre}, J., {Ahmed}, Z., {et~al.} 2019, \jcap, 2019, 056

\bibitem[{{Adri{\'a}n-Mart{\'\i}nez} {et~al.}(2016){Adri{\'a}n-Mart{\'\i}nez},
  {Ageron}, {Aharonian}, {Aiello}, {Albert}, {Ameli}, {Anassontzis}, {Andre},
  {Androulakis}, {Anghinolfi}, {Anton}, {Ardid}, {Avgitas}, {Barbarino},
  {Barbarito}, {Baret}, {Barrios-Mart{\'\i}}, {Belhorma}, {Belias}, {Berbee},
  {van den Berg}, {Bertin}, {Beurthey}, {van Beveren}, {Beverini}, {Biagi},
  {Biagioni}, {Billault}, {Bond{\`\i}}, {Bormuth}, {Bouhadef}, {Bourlis},
  {Bourret}, {Boutonnet}, {Bouwhuis}, {Bozza}, {Bruijn}, {Brunner}, {Buis},
  {Busto}, {Cacopardo}, {Caillat}, {Calamai}, {Calvo}, {Capone}, {Caramete},
  {Cecchini}, {Celli}, {Champion}, {Cherkaoui El Moursli}, {Cherubini},
  {Chiarusi}, {Circella}, {Classen}, {Cocimano}, {Coelho}, {Coleiro},
  {Colonges}, {Coniglione}, {Cordelli}, {Cosquer}, {Coyle}, {Creusot},
  {Cuttone}, {D'Amico}, {De Bonis}, {De Rosa}, {De Sio}, {Di Capua}, {Di
  Palma}, {D{\'\i}az Garc{\'\i}a}, {Distefano}, {Donzaud}, {Dornic},
  {Dorosti-Hasankiadeh}, {Drakopoulou}, {Drouhin}, {Drury}, {Durocher},
  {Eberl}, {Eichie}, {van Eijk}, {El Bojaddaini}, {El Khayati}, {Elsaesser},
  {Enzenh{\"o}fer}, {Fassi}, {Favali}, {Fermani}, {Ferrara}, {Filippidis},
  {Frascadore}, {Fusco}, {Gal}, {Galat{\`a}}, {Garufi}, {Gay}, {Gebyehu},
  {Giordano}, {Gizani}, {Gracia}, {Graf}, {Gr{\'e}goire}, {Grella}, {Habel},
  {Hallmann}, {van Haren}, {Harissopulos}, {Heid}, {Heijboer}, {Heine},
  {Henry}, {Hern{\'a}ndez-Rey}, {Hevinga}, {Hofest{\"a}dt}, {Hugon},
  {Illuminati}, {James}, {Jansweijer}, {Jongen}, {de Jong}, {Kadler},
  {Kalekin}, {Kappes}, {Katz}, {Keller}, {Kieft}, {Kie{\ss}ling}, {Koffeman},
  {Kooijman}, {Kouchner}, {Kulikovskiy}, {Lahmann}, {Lamare}, {Leisos},
  {Leonora}, {Clark}, {Liolios}, {Llorens Alvarez}, {Lo Presti}, {L{\"o}hner},
  {Lonardo}, {Lotze}, {Loucatos}, {Maccioni}, {Mannheim}, {Margiotta},
  {Marinelli}, {Mari{\c{s}}}, {Markou}, {Mart{\'\i}nez-Mora}, {Martini},
  {Mele}, {Melis}, {Michael}, {Migliozzi}, {Migneco}, {Mijakowski}, {Miraglia},
  {Mollo}, {Mongelli}, {Morganti}, {Moussa}, {Musico}, {Musumeci}, {Navas},
  {Nicolau}, {Olcina}, {Olivetto}, {Orlando}, {Papaikonomou}, {Papaleo},
  {P{\u{a}}v{\u{a}}la{\c{s}}}, {Peek}, {Pellegrino}, {Perrina}, {Pfutzner},
  {Piattelli}, {Pikounis}, {Poma}, {Popa}, {Pradier}, {Pratolongo},
  {P{\"u}hlhofer}, {Pulvirenti}, {Quinn}, {Racca}, {Raffaelli}, {Randazzo},
  {Rapidis}, {Razis}, {Real}, {Resvanis}, {Reubelt}, {Riccobene}, {Rossi},
  {Rovelli}, {Salda{\~n}a}, {Salvadori}, {Samtleben}, {S{\'a}nchez
  Garc{\'\i}a}, {S{\'a}nchez Losa}, {Sanguineti}, {Santangelo}, {Santonocito},
  {Sapienza}, {Schimmel}, {Schmelling}, {Sciacca}, {Sedita}, {Seitz}, {Sgura},
  {Simeone}, {Siotis}, {Sipala}, {Spisso}, {Spurio}, {Stavropoulos},
  {Steijger}, {Stellacci}, {Stransky}, {Taiuti}, {Tayalati}, {T{\'e}zier},
  {Theraube}, {Thompson}, {Timmer}, {T{\"o}nnis}, {Trasatti}, {Trovato},
  {Tsirigotis}, {Tzamarias}, {Tzamariudaki}, {Vallage}, {Van Elewyck},
  {Vermeulen}, {Vicini}, {Viola}, {Vivolo}, {Volkert}, {Voulgaris}, {Wiggers},
  {Wilms}, {de Wolf}, {Zachariadou}, {Zornoza}, \&
  {Z{\'u}{\~n}iga}}]{Km3net2016}
{Adri{\'a}n-Mart{\'\i}nez}, S., {Ageron}, M., {Aharonian}, F., {et~al.} 2016,
  Journal of Physics G Nuclear Physics, 43, 084001

\bibitem[{{Aiello} {et~al.}(2019){Aiello}, {Akrame}, {Ameli}, {Anassontzis},
  {Andre}, {Androulakis}, {Anghinolfi}, {Anton}, {Ardid}, {Aublin}, {Avgitas},
  {Bagatelas}, {Barbarino}, {Baret}, {Barrios-Mart{\'\i}}, {Belias}, {Berbee},
  {van den Berg}, {Bertin}, {Biagi}, {Biagioni}, {Biernoth}, {Boumaaza},
  {Bourret}, {Bouta}, {Bouwhuis}, {Bozza}, {Br{\^a}nza{\c{s}}}, {Bruchner},
  {Bruijn}, {Brunner}, {Buis}, {Buompane}, {Busto}, {Calvo}, {Capone}, {Celli},
  {Chabab}, {Chau}, {Cherubini}, {Chiarella}, {Chiarusi}, {Circella},
  {Cocimano}, {Coelho}, {Coleiro}, {Molla}, {Coniglione}, {Coyle}, {Creusot},
  {Cuttone}, {D'Onofrio}, {Dallier}, {De Sio}, {Di Palma}, {D{\'\i}az},
  {Diego-Tortosa}, {Distefano}, {Domi}, {Don{\`a}}, {Donzaud}, {Dornic},
  {D{\"o}rr}, {Durocher}, {Eberl}, {van Eijk}, {El Bojaddaini}, {Eljarrari},
  {Elsaesser}, {Enzenh{\"o}fer}, {Fermani}, {Ferrara}, {Filipovi{\'c}},
  {Fusco}, {Gal}, {Garcia}, {Garufi}, {Gialanella}, {Giorgio}, {Giuliante},
  {Gozzini}, {Gracia}, {Graf}, {Grasso}, {Gr{\'e}goire}, {Grella}, {Hallmann},
  {Hamdaoui}, {van Haren}, {Heid}, {Heijboer}, {Hekalo}, {Hern{\'a}ndez-Rey},
  {Hofest{\"a}dt}, {Illuminati}, {James}, {Jongen}, {de Jong}, {de Jong},
  {Kadler}, {Kalaczy{\'n}ski}, {Kalekin}, {Katz}, {Khan Chowdhury},
  {Kie{\ss}ling}, {Koffeman}, {Kooijman}, {Kouchner}, {Kreter}, {Kulikovskiy},
  {Kunhikannan Kannichankandy}, {Lahmann}, {Larosa}, {Le Breton}, {Leone},
  {Leonora}, {Levi}, {Lincetto}, {Lonardo}, {Longhitano}, {Lopez Coto},
  {Lotze}, {Maderer}, {Maggi}, {Ma{\'n}czak}, {Mannheim}, {Margiotta},
  {Marinelli}, {Markou}, {Martin}, {Mart{\'\i}nez-Mora}, {Martini},
  {Marzaioli}, {Mele}, {Melis}, {Migliozzi}, {Migneco}, {Mijakowski},
  {Miranda}, {Mollo}, {Morganti}, {Moser}, {Moussa}, {Muller}, {Musumeci},
  {Nauta}, {Navas}, {Nicolau}, {Nielsen}, {{\'O} Fearraigh}, {Organokov},
  {Orlando}, {Ottonello}, {Panagopoulos}, {Papalashvili}, {Papaleo},
  {P{\u{a}}v{\u{a}}la{\c{s}}}, {Pellegrino}, {Perrin-Terrin}, {Piattelli},
  {Pikounis}, {Pisanti}, {Poir{\'e}}, {Polydefki}, {Popa}, {Post}, {Pradier},
  {P{\"u}hlhofer}, {Pulvirenti}, {Quinn}, {Raffaelli}, {Randazzo}, {Razzaque},
  {Real}, {Resvanis}, {Reubelt}, {Riccobene}, {Richer}, {Rigalleau}, {Rovelli},
  {Saffer}, {Salvadori}, {Samtleben}, {S{\'a}nchez Losa}, {Sanguineti},
  {Santangelo}, {Santonocito}, {Sapienza}, {Schumann}, {Sciacca}, {Seneca},
  {Sgura}, {Shanidze}, {Sharma}, {Simeone}, {Sinopoulou}, {Spisso}, {Spurio},
  {Stavropoulos}, {Steijger}, {Stellacci}, {Strandberg}, {Stransky},
  {St{\"u}ven}, {Taiuti}, {Tatone}, {Tayalati}, {Tenllado}, {Thakore},
  {Trovato}, {Tzamariudaki}, {Tzanetatos}, {Van Elewyck}, {Versari}, {Viola},
  {Vivolo}, {Wilms}, {de Wolf}, {Zaborov}, {Zornoza}, {Z{\'u}{\~n}iga}, \&
  {KM3NeT Collaboration}}]{Aiello2019}
{Aiello}, S., {Akrame}, S.~E., {Ameli}, F., {et~al.} 2019, Astroparticle
  Physics, 111, 100

\bibitem[{{Anchordoqui} {et~al.}(2005){Anchordoqui}, {Goldberg}, {Halzen}, \&
  {Weiler}}]{2005PhLB..621...18A}
{Anchordoqui}, L.~A., {Goldberg}, H., {Halzen}, F., \& {Weiler}, T.~J. 2005,
  Physics Letters B, 621, 18

\bibitem[{{Ansoldi} {et~al.}(2018){Ansoldi}, {Antonelli}, {Arcaro}, {Baack},
  {Babi{\'c}}, {Banerjee}, {Bangale}, {Barres de Almeida}, {Barrio}, {Becerra
  Gonz{\'a}lez}, {Bednarek}, {Bernardini}, {Berse}, {Berti}, {Besenrieder},
  {Bhattacharyya}, {Bigongiari}, {Biland}, {Blanch}, {Bonnoli}, {Carosi},
  {Ceribella}, {Chatterjee}, {Colak}, {Colin}, {Colombo}, {Contreras},
  {Cortina}, {Covino}, {Cumani}, {D'Elia}, {Da Vela}, {Dazzi}, {De Angelis},
  {De Lotto}, {Delfino}, {Delgado}, {Di Pierro}, {Dom{\'\i}nguez}, {Dominis
  Prester}, {Dorner}, {Doro}, {Einecke}, {Elsaesser}, {Fallah Ramazani},
  {Fattorini}, {Fern{\'a}ndez-Barral}, {Ferrara}, {Fidalgo}, {Foffano},
  {Fonseca}, {Font}, {Fruck}, {Gallozzi}, {Garc{\'\i}a L{\'o}pez},
  {Garczarczyk}, {Gaug}, {Giammaria}, {Godinovi{\'c}}, {Guberman}, {Hadasch},
  {Hahn}, {Hassan}, {Hayashida}, {Herrera}, {Hoang}, {Hrupec}, {Inoue},
  {Ishio}, {Iwamura}, {Konno}, {Kubo}, {Kushida}, {Lamastra}, {Lelas}, {Leone},
  {Lindfors}, {Lombardi}, {Longo}, {L{\'o}pez}, {Maggio}, {Majumdar},
  {Makariev}, {Maneva}, {Manganaro}, {Mannheim}, {Maraschi}, {Mariotti},
  {Mart{\'\i}nez}, {Masuda}, {Mazin}, {Mielke}, {Minev}, {Miranda}, {Mirzoyan},
  {Moralejo}, {Moreno}, {Moretti}, {Neustroev}, {Niedzwiecki}, {Nievas
  Rosillo}, {Nigro}, {Nilsson}, {Ninci}, {Nishijima}, {Noda}, {Nogu{\'e}s},
  {Paiano}, {Palacio}, {Paneque}, {Paoletti}, {Paredes}, {Pedaletti},
  {Pe{\~n}il}, {Peresano}, {Persic}, {Pfrang}, {Prada Moroni}, {Prandini},
  {Puljak}, {Garcia}, {Rhode}, {Rib{\'o}}, {Rico}, {Righi}, {Rugliancich},
  {Saha}, {Saito}, {Satalecka}, {Schweizer}, {Sitarek}, {{\v{S}}nidari{\'c}},
  {Sobczynska}, {Stamerra}, {Strzys}, {Suri{\'c}}, {Tavecchio}, {Temnikov},
  {Terzi{\'c}}, {Teshima}, {Torres-Alb{\'a}}, {Tsujimoto}, {Vanzo}, {Vazquez
  Acosta}, {Vovk}, {Ward}, {Will}, {Zari{\'c}}, \&
  {Cerruti}}]{2018ApJ...863L..10A}
{Ansoldi}, S., {Antonelli}, L.~A., {Arcaro}, C., {et~al.} 2018, \apjl, 863, L10

\bibitem[{{Belolaptikov} {et~al.}(2022){Belolaptikov}, {Dzhilkibaev},
  {Allakhverdyan}, {Avrorin}, {Avrorin}, {Aynutdinov}, {Bannasch},
  {Barda{\v{c}}ov{\'a}}, {Belolaptikov}, {Borina}, {Brudanin}, {Budnev}, {Dik},
  {Domogatsky}, {Doroshenko}, {Dvornick{\'y}}, {Dyachok}, {Eckerov{\'a}},
  {Elzhov}, {Fajt}, {Fialkovsky}, {Gafarov}, {Golubkov}, {Gorshkov}, {Gress},
  {Katulin}, {Kebkal}, {Kebkal}, {Khramov}, {Kolbin}, {Konischev},
  {Kopa{\'n}ski}, {Korobchenko}, {Koshechkin}, {Kozhin}, {Kruglov}, {Kryukov},
  {Kulepov}, {Malecki}, {Malyshkin}, {Milenin}, {Mirgazov}, {Naumov}, {Nazari},
  {Noga}, {Petukhov}, {Pliskovsky}, {Rozanov}, {Rushay}, {Ryabov}, {Safronov},
  {Shaybonov}, {Shelepov}, {{\v{S}}imkovic}, {Sirenko}, {Skurikhin},
  {Solovjev}, {Sorokovikov}, {{\v{S}}tekl}, {Stromakov}, {Sushenok},
  {Suvorova}, {Tabolenko}, {Tarashansky}, {Yablokova}, {Yakovlev}, \&
  {Zaborov}}]{Belolaptikov2021}
{Belolaptikov}, I., {Dzhilkibaev}, Z.~A.~M., {Allakhverdyan}, V.~A., {et~al.}
  2022, in 37th International Cosmic Ray Conference. 12-23 July 2021. Berlin, 2

\bibitem[{{Blaufuss} {et~al.}(2019){Blaufuss}, {Kintscher}, {Lu}, \&
  {Tung}}]{Blaufuss2019}
{Blaufuss}, E., {Kintscher}, T., {Lu}, L., \& {Tung}, C.~F. 2019, in
  International Cosmic Ray Conference, Vol.~36, 36th International Cosmic Ray
  Conference (ICRC2019), 1021

\bibitem[{{B{\"o}ttcher}(2019)}]{2019Galax...7...20B}
{B{\"o}ttcher}, M. 2019, Galaxies, 7, 20

\bibitem[{{Capel} {et~al.}(2022){Capel}, {Burgess}, {Mortlock}, \&
  {Padovani}}]{Capel2022}
{Capel}, F., {Burgess}, J.~M., {Mortlock}, D.~J., \& {Padovani}, P. 2022, arXiv
  e-prints, arXiv:2201.05633

\bibitem[{{Cendes} {et~al.}(2021){Cendes}, {Alexander}, {Berger}, {Eftekhari},
  {Williams}, \& {Chornock}}]{Cendes2021}
{Cendes}, Y., {Alexander}, K.~D., {Berger}, E., {et~al.} 2021, \apj, 919, 127

\bibitem[{{Cerruti} {et~al.}(2019){Cerruti}, {Zech}, {Boisson}, {Emery},
  {Inoue}, \& {Lenain}}]{2019MNRAS.483L..12C}
{Cerruti}, M., {Zech}, A., {Boisson}, C., {et~al.} 2019, \mnras, 483, L12

\bibitem[{{Creque-Sarbinowski} {et~al.}(2022){Creque-Sarbinowski},
  {Kamionkowski}, \& {Zhou}}]{Creque-Sarbinowski2022}
{Creque-Sarbinowski}, C., {Kamionkowski}, M., \& {Zhou}, B. 2022, \prd, 105,
  123035

\bibitem[{{Giommi} {et~al.}(2020){Giommi}, {Glauch}, {Padovani}, {Resconi},
  {Turcati}, \& {Chang}}]{2020MNRAS.497..865G}
{Giommi}, P., {Glauch}, T., {Padovani}, P., {et~al.} 2020, \mnras, 497, 865

\bibitem[{{Globus} {et~al.}(2017){Globus}, {Allard}, {Parizot}, \&
  {Piran}}]{Globus2017}
{Globus}, N., {Allard}, D., {Parizot}, E., \& {Piran}, T. 2017, \apjl, 839, L22

\bibitem[{{Halzen} {et~al.}(2019){Halzen}, {Kheirandish}, {Weisgarber}, \&
  {Wakely}}]{2019ApJ...874L...9H}
{Halzen}, F., {Kheirandish}, A., {Weisgarber}, T., \& {Wakely}, S.~P. 2019,
  \apjl, 874, L9

\bibitem[{{Halzen} \& {Zas}(1997)}]{Halzen1997}
{Halzen}, F. \& {Zas}, E. 1997, \apj, 488, 669

\bibitem[{{Hovatta} {et~al.}(2021){Hovatta}, {Lindfors}, {Kiehlmann},
  {Max-Moerbeck}, {Hodges}, {Liodakis}, {L{\"a}hteem{\"a}ki}, {Pearson},
  {Readhead}, {Reeves}, {Suutarinen}, {Tammi}, \&
  {Tornikoski}}]{2021A&A...650A..83H}
{Hovatta}, T., {Lindfors}, E., {Kiehlmann}, S., {et~al.} 2021, \aap, 650, A83

\bibitem[{{Huber}(2019)}]{2019ICRC...36..916H}
{Huber}, M. 2019, in International Cosmic Ray Conference, Vol.~36, 36th
  International Cosmic Ray Conference (ICRC2019), 916

\bibitem[{{IceCube Collaboration}(2013)}]{2013Sci...342E...1I}
{IceCube Collaboration}. 2013, Science, 342, 1242856

\bibitem[{{IceCube Collaboration} {et~al.}(2018){IceCube Collaboration},
  {Aartsen}, {Ackermann}, {Adams}, {Aguilar}, {Ahlers}, {Ahrens}, {Al Samarai},
  {Altmann}, {Andeen}, {Anderson}, {Ansseau}, {Anton}, {Arg{\"u}elles},
  {Auffenberg}, {Axani}, {Bagherpour}, {Bai}, {Barron}, {Barwick}, {Baum},
  {Bay}, {Beatty}, {Becker Tjus}, {Becker}, {BenZvi}, {Berley}, {Bernardini},
  {Besson}, {Binder}, {Bindig}, {Blaufuss}, {Blot}, {Bohm}, {B{\"o}rner},
  {Bos}, {B{\"o}ser}, {Botner}, {Bourbeau}, {Bourbeau}, {Bradascio}, {Braun},
  {Brenzke}, {Bretz}, {Bron}, {Brostean-Kaiser}, {Burgman}, {Busse}, {Carver},
  {Cheung}, {Chirkin}, {Christov}, {Clark}, {Classen}, {Coenders}, {Collin},
  {Conrad}, {Coppin}, {Correa}, {Cowen}, {Cross}, {Dave}, {Day}, {de
  Andr{\'e}}, {De Clercq}, {DeLaunay}, {Dembinski}, {De Ridder}, {Desiati}, {de
  Vries}, {de Wasseige}, {de With}, {DeYoung}, {D{\'\i}az-V{\'e}lez}, {di
  Lorenzo}, {Dujmovic}, {Dumm}, {Dunkman}, {Dvorak}, {Eberhardt}, {Ehrhardt},
  {Eichmann}, {Eller}, {Evenson}, {Fahey}, {Fazely}, {Felde}, {Filimonov},
  {Finley}, {Flis}, {Franckowiak}, {Friedman}, {Fritz}, {Gaisser}, {Gallagher},
  {Gerhardt}, {Ghorbani}, {Glauch}, {Gl{\"u}senkamp}, {Goldschmidt},
  {Gonzalez}, {Grant}, {Griffith}, {Haack}, {Hallgren}, {Halzen}, {Hanson},
  {Hebecker}, {Heereman}, {Helbing}, {Hellauer}, {Hickford}, {Hignight},
  {Hill}, {Hoffman}, {Hoffmann}, {Hoinka}, {Hokanson-Fasig}, {Hoshina},
  {Huang}, {Huber}, {Hultqvist}, {H{\"u}nnefeld}, {Hussain}, {In}, {Iovine},
  {Ishihara}, {Jacobi}, {Japaridze}, {Jeong}, {Jero}, {Jones}, {Kalaczynski},
  {Kang}, {Kappes}, {Kappesser}, {Karg}, {Karle}, {Katz}, {Kauer}, {Keivani},
  {Kelley}, {Kheirandish}, {Kim}, {Kim}, {Kintscher}, {Kiryluk}, {Kittler},
  {Klein}, {Koirala}, {Kolanoski}, {K{\"o}pke}, {Kopper}, {Kopper},
  {Koschinsky}, {Koskinen}, {Kowalski}, {Krings}, {Kroll}, {Kr{\"u}ckl},
  {Kunwar}, {Kurahashi}, {Kuwabara}, {Kyriacou}, {Labare}, {Lanfranchi},
  {Larson}, {Lauber}, {Leonard}, {Lesiak-Bzdak}, {Leuermann}, {Liu}, {Lozano
  Mariscal}, {Lu}, {L{\"u}nemann}, {Luszczak}, {Madsen}, {Maggi}, {Mahn},
  {Mancina}, {Maruyama}, {Mase}, {Maunu}, {Meagher}, {Medici}, {Meier},
  {Menne}, {Merino}, {Meures}, {Miarecki}, {Micallef}, {Moment{\'e}},
  {Montaruli}, {Moore}, {S}, {Morse}, {Moulai}, {Nahnhauer}, {Nakarmi},
  {Naumann}, {Neer}, {Niederhausen}, {Nowicki}, {Nygren}, {Obertacke Pollmann},
  {Olivas}, {O'Murchadha}, {O'Sullivan}, {Palczewski}, {Pandya}, {Pankova},
  {Peiffer}, {Pepper}, {P{\'e}rez de los Heros}, {Pieloth}, {Pinat}, {Plum},
  {Price}, {Przybylski}, {Raab}, {R{\"a}del}, {Rameez}, {Rauch}, {Rawlins},
  {Rea}, {Reimann}, {Relethford}, {Relich}, {Resconi}, {Rhode}, {Richman},
  {Robertson}, {Rongen}, {Rott}, {Ruhe}, {Ryckbosch}, {Rysewyk}, {Safa},
  {S{\"a}lzer}, {Sanchez Herrera}, {Sandrock}, {Sandroos}, {Santander},
  {Sarkar}, {Sarkar}, {Satalecka}, {Schlunder}, {Schmidt}, {Schneider},
  {Schoenen}, {Sch{\"o}neberg}, {Schumacher}, {Sclafani}, {Seckel},
  {Seunarine}, {Soedingrekso}, {Soldin}, {Song}, {Spiczak}, {Spiering},
  {Stachurska}, {Stamatikos}, {Stanev}, {Stasik}, {Stein}, {Stettner},
  {Steuer}, {Stezelberger}, {Stokstad}, {St{\"o}{\ss}l}, {Strotjohann},
  {Stuttard}, {Sullivan}, {Sutherland}, {Taboada}, {Tatar}, {Tenholt},
  {Ter-Antonyan}, {Terliuk}, {Tilav}, {Toale}, {Tobin}, {Toennis}, {Toscano},
  {Tosi}, {Tselengidou}, {Tung}, {Turcati}, {Turley}, {Ty}, {Unger}, {Usner},
  {Vandenbroucke}, {Van Driessche}, {van Eijk}, {van Eijndhoven}, {Vanheule},
  {van Santen}, {Vogel}, {Vraeghe}, {Walck}, {Wallace}, {Wallraff}, {Wandler},
  {Wandkowsky}, {Waza}, {Weaver}, {Weiss}, {Wendt}, {Werthebach}, {Westerhoff},
  {Whelan}, {Whitehorn}, {Wiebe}, {Wiebusch}, {Wille}, {Williams}, {Wills},
  {Wolf}, {Wood}, {Wood}, {Woschnagg}, {Xu}, {Xu}, {Xu}, {Yanez}, {Yodh},
  {Yoshida}, {Yuan}, {Fermi-LAT Collaboration}, {Abdollahi}, {Ajello},
  {Angioni}, {Baldini}, {Ballet}, {Barbiellini}, {Bastieri}, {Bechtol},
  {Bellazzini}, {Berenji}, {Bissaldi}, {Blandford}, {Bonino}, {Bottacini},
  {Bregeon}, {Bruel}, {Buehler}, {Burnett}, {Burns}, {Buson}, {Cameron},
  {Caputo}, {Caraveo}, {Cavazzuti}, {Charles}, {Chen}, {Cheung}, {Chiang},
  {Chiaro}, {Ciprini}, {Cohen-Tanugi}, {Conrad}, {Costantin}, {Cutini},
  {D'Ammando}, {de Palma}, {Digel}, {Di Lalla}, {Di Mauro}, {Di Venere},
  {Dom{\'\i}nguez}, {Favuzzi}, {Franckowiak}, {Fukazawa}, {Funk}, {Fusco},
  {Gargano}, {Gasparrini}, {Giglietto}, {Giomi}, {Giommi}, {Giordano},
  {Giroletti}, {Glanzman}, {Green}, {Grenier}, {Grondin}, {Guiriec}, {Harding},
  {Hayashida}, {Hays}, {Hewitt}, {Horan}, {J{\'o}hannesson}, {Kadler},
  {Kensei}, {Kocevski}, {Krauss}, {Kreter}, {Kuss}, {La Mura}, {Larsson},
  {Latronico}, {Lemoine-Goumard}, {Li}, {Longo}, {Loparco}, {Lovellette},
  {Lubrano}, {Magill}, {Maldera}, {Malyshev}, {Manfreda}, {Mazziotta},
  {McEnery}, {Meyer}, {Michelson}, {Mizuno}, {Monzani}, {Morselli},
  {Moskalenko}, {Negro}, {Nuss}, {Ojha}, {Omodei}, {Orienti}, {Orlando},
  {Palatiello}, {Paliya}, {Perkins}, {Persic}, {Pesce-Rollins}, {Piron},
  {Porter}, {Principe}, {Rain{\`o}}, {Rando}, {Rani}, {Razzano}, {Razzaque},
  {Reimer}, {Reimer}, {Renault-Tinacci}, {Ritz}, {Rochester}, {Saz Parkinson},
  {Sgr{\`o}}, {Siskind}, {Spandre}, {Spinelli}, {Suson}, {Tajima}, {Takahashi},
  {Tanaka}, {Thayer}, {Thompson}, {Tibaldo}, {Torres}, {Torresi}, {Tosti},
  {Troja}, {Valverde}, {Vianello}, {Vogel}, {Wood}, {Wood}, {Zaharijas}, {MAGIC
  Collaboration}, {Ahnen}, {Ansoldi}, {Antonelli}, {Arcaro}, {Baack},
  {Babi{\'c}}, {Banerjee}, {Bangale}, {Barres de Almeida}, {Barrio}, {Becerra
  Gonz{\'a}lez}, {Bednarek}, {Bernardini}, {Berti}, {Bhattacharyya}, {Biland},
  {Blanch}, {Bonnoli}, {Carosi}, {Carosi}, {Ceribella}, {Chatterjee}, {Colak},
  {Colin}, {Colombo}, {Contreras}, {Cortina}, {Covino}, {Cumani}, {Da Vela},
  {Dazzi}, {De Angelis}, {De Lotto}, {Delfino}, {Delgado}, {Di Pierro},
  {Dom{\'\i}nguez}, {Dominis Prester}, {Dorner}, {Doro}, {Einecke},
  {Elsaesser}, {Fallah Ramazani}, {Fern{\'a}ndez-Barral}, {Fidalgo}, {Foffano},
  {Pfrang}, {Fonseca}, {Font}, {Franceschini}, {Fruck}, {Galindo}, {Gallozzi},
  {Garc{\'\i}a L{\'o}pez}, {Garczarczyk}, {Gaug}, {Giammaria}, {Godinovi{\'c}},
  {Gora}, {Guberman}, {Hadasch}, {Hahn}, {Hassan}, {Hayashida}, {Herrera},
  {Hose}, {Hrupec}, {Inoue}, {Ishio}, {Konno}, {Kubo}, {Kushida}, {Lelas},
  {Lindfors}, {Lombardi}, {Longo}, {L{\'o}pez}, {Maggio}, {Majumdar},
  {Makariev}, {Maneva}, {Manganaro}, {Mannheim}, {Maraschi}, {Mariotti},
  {Mart{\'\i}nez}, {Masuda}, {Mazin}, {Minev}, {M}, {Mirzoyan}, {Moralejo},
  {Moreno}, {Moretti}, {Nagayoshi}, {Neustroev}, {Niedzwiecki}, {Nievas
  Rosillo}, {Nigro}, {Nilsson}, {Ninci}, {Nishijima}, {Noda}, {Nogu{\'e}s},
  {Paiano}, {Palacio}, {Paneque}, {Paoletti}, {Paredes}, {Pedaletti},
  {Peresano}, {Persic}, {Prada Moroni}, {Prandini}, {Puljak}, {Rodriguez
  Garcia}, {Reichardt}, {Rhode}, {Rib{\'o}}, {Rico}, {Righi}, {Rugliancich},
  {Saito}, {Satalecka}, {Schweizer}, {Sitarek}, {{\v{S}}nidari{\'c}},
  {Sobczynska}, {Stamerra}, {Strzys}, {Suri{\'c}}, {Takahashi}, {Tavecchio},
  {Temnikov}, {Terzi{\'c}}, {Teshima}, {Torres-Alb{\`a}}, {Treves},
  {Tsujimoto}, {Vanzo}, {Vazquez Acosta}, {Vovk}, {Ward}, {Will}, {S},
  {Zari{\'c}}, {AGILE Team}, {Lucarelli}, {Tavani}, {Piano}, {Donnarumma},
  {Pittori}, {Verrecchia}, {Barbiellini}, {Bulgarelli}, {Caraveo}, {Cattaneo},
  {Colafrancesco}, {Costa}, {Di Cocco}, {Ferrari}, {Gianotti}, {Giuliani},
  {Lipari}, {Mereghetti}, {Morselli}, {Pacciani}, {Paoletti}, {Parmiggiani},
  {Pellizzoni}, {Picozza}, {Pilia}, {Rappoldi}, {Trois}, {Vercellone},
  {Vittorini}, {ASAS-SN Team}, {Stanek}, {Kochanek}, {Beacom}, {Thompson},
  {Holoien}, {Dong}, {Prieto}, {Shappee}, {Holmbo}, {HAWC Collaboration},
  {Abeysekara}, {Albert}, {Alfaro}, {Alvarez}, {Arceo},
  {Arteaga-Vel{\'a}zquez}, {Avila Rojas}, {Ayala Solares}, {Becerril},
  {Belmont-Moreno}, {Bernal}, {Caballero-Mora}, {Capistr{\'a}n},
  {Carrami{\~n}ana}, {Casanova}, {Castillo}, {Cotti}, {Cotzomi}, {Couti{\~n}o
  de Le{\'o}n}, {De Le{\'o}n}, {De la Fuente}, {Diaz Hernandez}, {Dichiara},
  {Dingus}, {DuVernois}, {D{\'\i}az-V{\'e}lez}, {Ellsworth}, {Engel},
  {Fiorino}, {Fleischhack}, {Fraija}, {Garc{\'\i}a-Gonz{\'a}lez}, {Garfias},
  {Gonz{\'a}lez Mu{\~n}oz}, {Gonz{\'a}lez}, {Goodman}, {Hampel-Arias},
  {Harding}, {Hernand ez}, {Hona}, {Hueyotl-Zahuantitla}, {Hui},
  {H{\"u}ntemeyer}, {Iriarte}, {Jardin-Blicq}, {Joshi}, {Kaufmann}, {Kunde},
  {Lara}, {Lauer}, {Lee}, {Lennarz}, {Le{\'o}n Vargas}, {Linnemann},
  {Longinotti}, {Luis-Raya}, {Luna-Garc{\'\i}a}, {Malone}, {Marinelli},
  {Martinez}, {Martinez-Castellanos}, {Mart{\'\i}nez-Castro},
  {Mart{\'\i}nez-Huerta}, {Matthews}, {Miranda-Romagnoli}, {Moreno},
  {Mostaf{\'a}}, {Nayerhoda}, {Nellen}, {Newbold}, {Nisa}, {Noriega-Papaqui},
  {Pelayo}, {Pretz}, {P{\'e}rez-P{\'e}rez}, {Ren}, {Rho}, {Rivi{\`e}re},
  {Rosa-Gonz{\'a}lez}, {Rosenberg}, {Ruiz-Velasco}, {Ruiz-Velasco}, {Salesa
  Greus}, {Sandoval}, {Schneider}, {Schoorlemmer}, {Sinnis}, {Smith},
  {Springer}, {Surajbali}, {Tibolla}, {Tollefson}, {Torres}, {Villase{\~n}or},
  {Weisgarber}, {Werner}, {Yapici}, {Gaurang}, {Zepeda}, {Zhou}, {{\'A}lvarez},
  {H.~E.~S.~S. Collaboration}, {Abdalla}, {Ang{\"u}ner}, {Armand}, {Backes},
  {Becherini}, {Berge}, {B{\"o}ttcher}, {Boisson}, {Bolmont}, {Bonnefoy},
  {Bordas}, {Brun}, {B{\"u}chele}, {Bulik}, {Caroff}, {Carosi}, {Casanova},
  {Cerruti}, {Chakraborty}, {Chandra}, {Chen}, {Colafrancesco}, {Davids},
  {Deil}, {Devin}, {Djannati-Ata{\"\i}}, {Egberts}, {Emery}, {Eschbach},
  {Fiasson}, {Fontaine}, {Funk}, {F{\"u}{\ss}ling}, {Gallant}, {Gat{\'e}},
  {Giavitto}, {Glawion}, {Glicenstein}, {Gottschall}, {Grondin}, {Haupt},
  {Henri}, {Hinton}, {Hoischen}, {Holch}, {Huber}, {Jamrozy}, {Jankowsky},
  {Jankowsky}, {Jouvin}, {Jung-Richardt}, {Kerszberg}, {Kh{\'e}lifi}, {King},
  {Klepser}, {Klu{\'z}niak}, {Komin}, {Kraus}, {Lefaucheur}, {Lemi{\`e}re},
  {Lemoine-Goumard}, {Lenain}, {Leser}, {Lohse}, {L{\'o}pez-Coto}, {Lorentz},
  {Lypova}, {Marandon}, {Guillem Mart{\'\i}-Devesa}, {Maurin}, {Mitchell},
  {Moderski}, {Mohamed}, {Mohrmann}, {Moulin}, {Murach}, {de Naurois},
  {Niederwanger}, {Niemiec}, {Oakes}, {O'Brien}, {Ohm}, {Ostrowski}, {Oya},
  {Panter}, {Parsons}, {Perennes}, {Piel}, {Pita}, {Poireau}, {Priyana Noel},
  {Prokoph}, {P{\"u}hlhofer}, {Quirrenbach}, {Raab}, {Rauth}, {Renaud},
  {Rieger}, {Rinchiuso}, {Romoli}, {Rowell}, {Rudak}, {Sasaki}, {Sanchez},
  {Schlickeiser}, {Sch{\"u}ssler}, {Schulz}, {Schwanke}, {Seglar-Arroyo},
  {Shafi}, {Simoni}, {Sol}, {Stegmann}, {Steppa}, {Tavernier}, {Taylor},
  {Tiziani}, {Trichard}, {Tsirou}, {van Eldik}, {van Rensburg}, {van Soelen},
  {Veh}, {Vincent}, {Voisin}, {Wagner}, {Wagner}, {Wierzcholska}, {Zanin},
  {Zdziarski}, {Zech}, {Ziegler}, {Zorn}, {{\.Z}ywucka}, {INTEGRAL Team},
  {Savchenko}, {Ferrigno}, {Bazzano}, {Diehl}, {Kuulkers}, {Laurent},
  {Mereghetti}, {Natalucci}, {Panessa}, {Rodi}, {Ubertini}, {Kanata}, Teams,
  {Morokuma}, {Ohta}, {Tanaka}, {Mori}, {Yamanaka}, {Kawabata}, {Utsumi},
  {Nakaoka}, {Kawabata}, {Nagashima}, {Yoshida}, {Matsuoka}, {Itoh}, {Kapteyn
  Team}, {Keel}, {Liverpool Telescope Team}, {Copperwheat}, {Steele},
  {Swift/NuSTAR Team}, {Cenko}, {Cowen}, {DeLaunay}, {Evans}, {Fox}, {Keivani},
  {Kennea}, {Marshall}, {Osborne}, {Santander}, {Tohuvavohu}, {Turley},
  {VERITAS Collaboration}, {Abeysekara}, {Archer}, {Benbow}, {Bird}, {Brill},
  {Brose}, {Buchovecky}, {Buckley}, {Bugaev}, {Christiansen}, {Connolly},
  {Cui}, {Daniel}, {Errando}, {Falcone}, {Feng}, {Finley}, {Fortson},
  {Furniss}, {Gueta}, {H{\"u}tten}, {Hervet}, {Hughes}, {Humensky}, {Johnson},
  {Kaaret}, {Kar}, {Kelley-Hoskins}, {Kertzman}, {Kieda}, {Krause},
  {Krennrich}, {Kumar}, {Lang}, {Lin}, {Maier}, {McArthur}, {Moriarty},
  {Mukherjee}, {Nieto}, {O'Brien}, {Ong}, {Otte}, {Park}, {Petrashyk}, {Pohl},
  {Popkow}, {Pueschel}, {Quinn}, {Ragan}, {Reynolds}, {Richards}, {Roache},
  {Rulten}, {Sadeh}, {Santander}, {Scott}, {Sembroski}, {Shahinyan}, {Sushch},
  {Tr{\'e}panier}, {Tyler}, {Vassiliev}, {Wakely}, {Weinstein}, {Wells},
  {Wilcox}, {Wilhelm}, {Williams}, {Zitzer}, {VLA/B Team}, {Tetarenko},
  {Kimball}, {Miller-Jones}, \& {Sivakoff}}]{2018Sci...361.1378I}
{IceCube Collaboration}, {Aartsen}, M.~G., {Ackermann}, M., {et~al.} 2018,
  Science, 361, eaat1378

\bibitem[{{Illuminati} {et~al.}(2021){Illuminati}, {Illuminati}, \& {ANTARES
  Collaboration}}]{Illuminati2021}
{Illuminati}, G., {Illuminati}, G., \& {ANTARES Collaboration}. 2021, Journal
  of Instrumentation, 16, C10005

\bibitem[{{Janssen} {et~al.}(2021){Janssen}, {Falcke}, {Kadler}, {Ros},
  {Wielgus}, {Akiyama}, {Balokovi{\'c}}, {Blackburn}, {Bouman}, {Chael},
  {Chan}, {Chatterjee}, {Davelaar}, {Edwards}, {Fromm}, {G{\'o}mez}, {Goddi},
  {Issaoun}, {Johnson}, {Kim}, {Koay}, {Krichbaum}, {Liu}, {Liuzzo}, {Markoff},
  {Markowitz}, {Marrone}, {Mizuno}, {M{\"u}ller}, {Ni}, {Pesce},
  {Ramakrishnan}, {Roelofs}, {Rygl}, {van Bemmel}, {Event Horizon Telescope
  Collaboration}, {Alberdi}, {Alef}, {Algaba}, {Anantua}, {Asada}, {Azulay},
  {Baczko}, {Ball}, {Ball}, {Barrett}, {Benson}, {Bintley}, {Bintley},
  {Blundell}, {Boland}, {Boland}, {Bower}, {Boyce}, {Bremer}, {Brinkerink},
  {Brissenden}, {Britzen}, {Broderick}, {Broguiere}, {Bronzwaer}, {Byun},
  {Carlstrom}, {Chatterjee}, {Chen}, {Chen}, {Chesler}, {Cho}, {Christian},
  {Conway}, {Cordes}, {Crawford}, {Crew}, {Cruz-Osorio}, {Cui}, {Cui}, {De
  Laurentis}, {Deane}, {Dempsey}, {Desvignes}, {Dexter}, {Doeleman}, {Eatough},
  {Farah}, {Farah}, {Fish}, {Fomalont}, {Ford}, {Fraga-Encinas}, {Friberg},
  {Friberg}, {Fuentes}, {Galison}, {Gammie}, {Garc{\'\i}a}, {Gelles}, {Gentaz},
  {Georgiev}, {Georgiev}, {Gold}, {Gold}, {G{\'o}mez-Ruiz}, {Gu}, {Gurwell},
  {Hada}, {Haggard}, {Hecht}, {Hesper}, {Himwich}, {Ho}, {Ho}, {Honma},
  {Huang}, {Huang}, {Hughes}, {Ikeda}, {Inoue}, {Inoue}, {James}, {Jannuzi},
  {Jannuzi}, {Jeter}, {Jiang}, {Jimenez-Rosales}, {Jimenez-Rosales}, {Jorstad},
  {Jung}, {Karami}, {Karuppusamy}, {Kawashima}, {Keating}, {Kettenis}, {Kim},
  {Kim}, {Kim}, {Kim}, {Kino}, {Kino}, {Kofuji}, {Koyama}, {Kramer}, {Kramer},
  {Kramer}, {Kuo}, {Lauer}, {Lee}, {Levis}, {Li}, {Li}, {Lindqvist}, {Lico},
  {Lindahl}, {Lindahl}, {Liu}, {Liu}, {Lo}, {Lobanov}, {Loinard}, {Lonsdale},
  {Lu}, {MacDonald}, {Mao}, {Marchili}, {Marchili}, {Marchili}, {Marscher},
  {Mart{\'\i}-Vidal}, {Matsushita}, {Matthews}, {Medeiros}, {Menten}, {Mizuno},
  {Mizuno}, {Moran}, {Moriyama}, {Moscibrodzka}, {Moscibrodzka}, {Musoke},
  {Mej{\'\i}as}, {Nagai}, {Nagar}, {Nakamura}, {Narayan}, {Narayanan},
  {Natarajan}, {Nathanail}, {Neilsen}, {Neri}, {Neri}, {Noutsos}, {Nowak},
  {Okino}, {Olivares}, {Ortiz-Le{\'o}n}, {Oyama}, {{\"O}zel}, {Palumbo},
  {Park}, {Patel}, {Pen}, {Pen}, {Pi{\'e}tu}, {Plambeck}, {PopStefanija},
  {Porth}, {P{\"o}tzl}, {Prather}, {Preciado-L{\'o}pez}, {Psaltis}, {Pu}, {Pu},
  {Rao}, {Rawlings}, {Raymond}, {Rezzolla}, {Ricarte}, {Ripperda}, {Ripperda},
  {Rogers}, {Rogers}, {Rose}, {Roshanineshat}, {Rottmann}, {Roy}, {Ruszczyk},
  {Ruszczyk}, {S{\'a}nchez}, {S{\'a}nchez-Arguelles}, {Sasada}, {Savolainen},
  {Schloerb}, {Schuster}, {Shao}, {Shen}, {Small}, {Sohn}, {SooHoo}, {Sun},
  {Tazaki}, {Tetarenko}, {Tiede}, {Tilanus}, {Titus}, {Torne}, {Trent},
  {Traianou}, {Trippe}, {van Bemmel}, {van Langevelde}, {van Rossum}, {Wagner},
  {Ward-Thompson}, {Wardle}, {Weintroub}, {Wex}, {Wharton}, {Wharton}, {Wong},
  {Wu}, {Yoon}, {Young}, {Young}, {Younsi}, {Yuan}, {Yuan}, {Zensus}, {Zhao},
  \& {Zhao}}]{Janssen2021}
{Janssen}, M., {Falcke}, H., {Kadler}, M., {et~al.} 2021, Nature Astronomy, 5,
  1017

\bibitem[{Keivani {et~al.}(2018)}]{keivani18}
Keivani, A. {et~al.} 2018, Astrophys. J., 864, 84

\bibitem[{{Kim} {et~al.}(2018){Kim}, {Krichbaum}, {Lu}, {Ros}, {Bach},
  {Bremer}, {de Vicente}, {Lindqvist}, \& {Zensus}}]{Kim2018}
{Kim}, J.~Y., {Krichbaum}, T.~P., {Lu}, R.~S., {et~al.} 2018, \aap, 616, A188

\bibitem[{{Kun} {et~al.}(2021){Kun}, {Bartos}, {Tjus}, {Biermann}, {Halzen}, \&
  {Mez{\H{o}}}}]{Kun2021}
{Kun}, E., {Bartos}, I., {Tjus}, J.~B., {et~al.} 2021, \apjl, 911, L18

\bibitem[{{Kun} {et~al.}(2019){Kun}, {Biermann}, \&
  {Gergely}}]{2019MNRAS.483L..42K}
{Kun}, E., {Biermann}, P.~L., \& {Gergely}, L.~{\'A}. 2019, \mnras, 483, L42

\bibitem[{{Liodakis} {et~al.}(2017){Liodakis}, {Pavlidou}, {Hovatta},
  {Max-Moerbeck}, {Pearson}, {Richards}, \& {Readhead}}]{Liodakis2017}
{Liodakis}, I., {Pavlidou}, V., {Hovatta}, T., {et~al.} 2017, \mnras, 467, 4565

\bibitem[{{Liodakis} \& {Petropoulou}(2020)}]{Liodakis2020}
{Liodakis}, I. \& {Petropoulou}, M. 2020, \apjl, 893, L20

\bibitem[{{Liodakis} {et~al.}(2018){Liodakis}, {Romani}, {Filippenko},
  {Kiehlmann}, {Max-Moerbeck}, {Readhead}, \& {Zheng}}]{Liodakis2018}
{Liodakis}, I., {Romani}, R.~W., {Filippenko}, A.~V., {et~al.} 2018, \mnras,
  480, 5517

\bibitem[{{Liodakis} {et~al.}(2019){Liodakis}, {Romani}, {Filippenko},
  {Kocevski}, \& {Zheng}}]{Liodakis2019}
{Liodakis}, I., {Romani}, R.~W., {Filippenko}, A.~V., {Kocevski}, D., \&
  {Zheng}, W. 2019, \apj, 880, 32

\bibitem[{{Liu} {et~al.}(2019){Liu}, {Wang}, {Xue}, {Taylor}, {Wang}, {Li}, \&
  {Yan}}]{2019PhRvD..99f3008L}
{Liu}, R.-Y., {Wang}, K., {Xue}, R., {et~al.} 2019, \prd, 99, 063008

\bibitem[{{Loeb} \& {Waxman}(2006)}]{Loeb2006}
{Loeb}, A. \& {Waxman}, E. 2006, \jcap, 2006, 003

\bibitem[{{LSST Science Collaboration} {et~al.}(2009){LSST Science
  Collaboration}, {Abell}, {Allison}, {Anderson}, {Andrew}, {Angel}, {Armus},
  {Arnett}, {Asztalos}, {Axelrod}, {Bailey}, {Ballantyne}, {Bankert},
  {Barkhouse}, {Barr}, {Barrientos}, {Barth}, {Bartlett}, {Becker}, {Becla},
  {Beers}, {Bernstein}, {Biswas}, {Blanton}, {Bloom}, {Bochanski}, {Boeshaar},
  {Borne}, {Bradac}, {Brandt}, {Bridge}, {Brown}, {Brunner}, {Bullock},
  {Burgasser}, {Burge}, {Burke}, {Cargile}, {Chandrasekharan}, {Chartas},
  {Chesley}, {Chu}, {Cinabro}, {Claire}, {Claver}, {Clowe}, {Connolly}, {Cook},
  {Cooke}, {Cooray}, {Covey}, {Culliton}, {de Jong}, {de Vries}, {Debattista},
  {Delgado}, {Dell'Antonio}, {Dhital}, {Di Stefano}, {Dickinson}, {Dilday},
  {Djorgovski}, {Dobler}, {Donalek}, {Dubois-Felsmann}, {Durech},
  {Eliasdottir}, {Eracleous}, {Eyer}, {Falco}, {Fan}, {Fassnacht}, {Ferguson},
  {Fernandez}, {Fields}, {Finkbeiner}, {Figueroa}, {Fox}, {Francke}, {Frank},
  {Frieman}, {Fromenteau}, {Furqan}, {Galaz}, {Gal-Yam}, {Garnavich},
  {Gawiser}, {Geary}, {Gee}, {Gibson}, {Gilmore}, {Grace}, {Green}, {Gressler},
  {Grillmair}, {Habib}, {Haggerty}, {Hamuy}, {Harris}, {Hawley}, {Heavens},
  {Hebb}, {Henry}, {Hileman}, {Hilton}, {Hoadley}, {Holberg}, {Holman},
  {Howell}, {Infante}, {Ivezic}, {Jacoby}, {Jain}, {R}, {Jedicke}, {Jee},
  {Garrett Jernigan}, {Jha}, {Johnston}, {Jones}, {Juric}, {Kaasalainen},
  {Styliani}, {Kafka}, {Kahn}, {Kaib}, {Kalirai}, {Kantor}, {Kasliwal},
  {Keeton}, {Kessler}, {Knezevic}, {Kowalski}, {Krabbendam}, {Krughoff},
  {Kulkarni}, {Kuhlman}, {Lacy}, {Lepine}, {Liang}, {Lien}, {Lira}, {Long},
  {Lorenz}, {Lotz}, {Lupton}, {Lutz}, {Macri}, {Mahabal}, {Mandelbaum},
  {Marshall}, {May}, {McGehee}, {Meadows}, {Meert}, {Milani}, {Miller},
  {Miller}, {Mills}, {Minniti}, {Monet}, {Mukadam}, {Nakar}, {Neill}, {Newman},
  {Nikolaev}, {Nordby}, {O'Connor}, {Oguri}, {Oliver}, {Olivier}, {Olsen},
  {Olsen}, {Olszewski}, {Oluseyi}, {Padilla}, {Parker}, {Pepper}, {Peterson},
  {Petry}, {Pinto}, {Pizagno}, {Popescu}, {Prsa}, {Radcka}, {Raddick},
  {Rasmussen}, {Rau}, {Rho}, {Rhoads}, {Richards}, {Ridgway}, {Robertson},
  {Roskar}, {Saha}, {Sarajedini}, {Scannapieco}, {Schalk}, {Schindler},
  {Schmidt}, {Schmidt}, {Schneider}, {Schumacher}, {Scranton}, {Sebag},
  {Seppala}, {Shemmer}, {Simon}, {Sivertz}, {Smith}, {Allyn Smith}, {Smith},
  {Spitz}, {Stanford}, {Stassun}, {Strader}, {Strauss}, {Stubbs}, {Sweeney},
  {Szalay}, {Szkody}, {Takada}, {Thorman}, {Trilling}, {Trimble}, {Tyson}, {Van
  Berg}, {Vanden Berk}, {VanderPlas}, {Verde}, {Vrsnak}, {Walkowicz},
  {Wandelt}, {Wang}, {Wang}, {Warner}, {Wechsler}, {West}, {Wiecha},
  {Williams}, {Willman}, {Wittman}, {Wolff}, {Wood-Vasey}, {Wozniak}, {Young},
  {Zentner}, \& {Zhan}}]{LSST2009}
{LSST Science Collaboration}, {Abell}, P.~A., {Allison}, J., {et~al.} 2009,
  arXiv e-prints, arXiv:0912.0201

\bibitem[{{Mannheim}(1995)}]{Mannheim1995}
{Mannheim}, K. 1995, Astroparticle Physics, 3, 295

\bibitem[{{Mastichiadis} \& {Petropoulou}(2021)}]{Mastichiadis2020}
{Mastichiadis}, A. \& {Petropoulou}, M. 2021, \apj, 906, 131

\bibitem[{{Max-Moerbeck} {et~al.}(2014){Max-Moerbeck}, {Hovatta}, {Richards},
  {King}, {Pearson}, {Readhead}, {Reeves}, {Shepherd}, {Stevenson},
  {Angelakis}, {Fuhrmann}, {Grainge}, {Pavlidou}, {Romani}, \&
  {Zensus}}]{maxmoerbeck2014}
{Max-Moerbeck}, W., {Hovatta}, T., {Richards}, J.~L., {et~al.} 2014, \mnras,
  445, 428

\bibitem[{{Mbarek} \& {Caprioli}(2021)}]{Mbarek2021}
{Mbarek}, R. \& {Caprioli}, D. 2021, \apj, 921, 85

\bibitem[{{Mirzoyan}(2017)}]{2017ATel10817....1M}
{Mirzoyan}, R. 2017, The Astronomer's Telegram, 10817, 1

\bibitem[{{Murase} {et~al.}(2016){Murase}, {Guetta}, \& {Ahlers}}]{Murase2016}
{Murase}, K., {Guetta}, D., \& {Ahlers}, M. 2016, \prl, 116, 071101

\bibitem[{Murase {et~al.}(2018)Murase, Oikonomou, \& Petropoulou}]{murase18}
Murase, K., Oikonomou, F., \& Petropoulou, M. 2018, Astrophys. J., 865, 124

\bibitem[{{Murase} {et~al.}(2018{\natexlab{a}}){Murase}, {Oikonomou}, \&
  {Petropoulou}}]{2018ApJ...865..124M}
{Murase}, K., {Oikonomou}, F., \& {Petropoulou}, M. 2018{\natexlab{a}}, \apj,
  865, 124

\bibitem[{{Murase} {et~al.}(2018{\natexlab{b}}){Murase}, {Oikonomou}, \&
  {Petropoulou}}]{Murase2018}
{Murase}, K., {Oikonomou}, F., \& {Petropoulou}, M. 2018{\natexlab{b}}, \apj,
  865, 124

\bibitem[{Neronov \& Semikoz(2002)}]{PhysRevD.66.123003}
Neronov, A.~Y. \& Semikoz, D.~V. 2002, Phys. Rev. D, 66, 123003

\bibitem[{{Nilsson} {et~al.}(2018){Nilsson}, {Lindfors}, {Takalo}, {Reinthal},
  {Berdyugin}, {Sillanp{\"a}{\"a}}, {Ciprini}, {Halkola}, {Hein{\"a}m{\"a}ki},
  {Hovatta}, {Kadenius}, {Nurmi}, {Ostorero}, {Pasanen}, {Rekola}, {Saarinen},
  {Sainio}, {Tuominen}, {Villforth}, {Vornanen}, \& {Zaprudin}}]{Nilsson2018}
{Nilsson}, K., {Lindfors}, E., {Takalo}, L.~O., {et~al.} 2018, \aap, 620, A185

\bibitem[{{Padovani} {et~al.}(2016){Padovani}, {Resconi}, {Giommi}, {Arsioli},
  \& {Chang}}]{2016MNRAS.457.3582P}
{Padovani}, P., {Resconi}, E., {Giommi}, P., {Arsioli}, B., \& {Chang}, Y.~L.
  2016, \mnras, 457, 3582

\bibitem[{{Petropoulou} {et~al.}(2020){Petropoulou}, {Murase}, {Santander},
  {Buson}, {Tohuvavohu}, {Kawamuro}, {Vasilopoulos}, {Negoro}, {Ueda},
  {Siegel}, {Keivani}, {Kawai}, {Mastichiadis}, \&
  {Dimitrakoudis}}]{2020ApJ...891..115P}
{Petropoulou}, M., {Murase}, K., {Santander}, M., {et~al.} 2020, \apj, 891, 115

\bibitem[{{Plavin} {et~al.}(2020){Plavin}, {Kovalev}, {Kovalev}, \&
  {Troitsky}}]{2020ApJ...894..101P}
{Plavin}, A., {Kovalev}, Y.~Y., {Kovalev}, Y.~A., \& {Troitsky}, S. 2020, \apj,
  894, 101

\bibitem[{{Plavin} {et~al.}(2021){Plavin}, {Kovalev}, {Kovalev}, \&
  {Troitsky}}]{2021ApJ...908..157P}
{Plavin}, A.~V., {Kovalev}, Y.~Y., {Kovalev}, Y.~A., \& {Troitsky}, S.~V. 2021,
  \apj, 908, 157

\bibitem[{{Reimer} {et~al.}(2019){Reimer}, {B{\"o}ttcher}, \&
  {Buson}}]{2019ApJ...881...46R}
{Reimer}, A., {B{\"o}ttcher}, M., \& {Buson}, S. 2019, \apj, 881, 46

\bibitem[{{Richards} {et~al.}(2011){Richards}, {Max-Moerbeck}, {Pavlidou},
  {King}, {Pearson}, {Readhead}, {Reeves}, {Shepherd}, {Stevenson},
  {Weintraub}, {Fuhrmann}, {Angelakis}, {Zensus}, {Healey}, {Romani}, {Shaw},
  {Grainge}, {Birkinshaw}, {Lancaster}, {Worrall}, {Taylor}, {Cotter}, \&
  {Bustos}}]{Richards2011}
{Richards}, J.~L., {Max-Moerbeck}, W., {Pavlidou}, V., {et~al.} 2011, \apjs,
  194, 29

\bibitem[{{Rodrigues} {et~al.}(2019){Rodrigues}, {Gao}, {Fedynitch},
  {Palladino}, \& {Winter}}]{2019ApJ...874L..29R}
{Rodrigues}, X., {Gao}, S., {Fedynitch}, A., {Palladino}, A., \& {Winter}, W.
  2019, \apjl, 874, L29

\bibitem[{{Sahakyan}(2018)}]{2018ApJ...866..109S}
{Sahakyan}, N. 2018, \apj, 866, 109

\bibitem[{{Scargle} {et~al.}(2013){Scargle}, {Norris}, {Jackson}, \&
  {Chiang}}]{Scargle2013}
{Scargle}, J.~D., {Norris}, J.~P., {Jackson}, B., \& {Chiang}, J. 2013, \apj,
  764, 167

\bibitem[{{Stecker} {et~al.}(1991){Stecker}, {Done}, {Salamon}, \&
  {Sommers}}]{Stecker1991}
{Stecker}, F.~W., {Done}, C., {Salamon}, M.~H., \& {Sommers}, P. 1991, \prl,
  66, 2697

\bibitem[{{Stein} {et~al.}(2021){Stein}, {Velzen}, {Kowalski}, {Franckowiak},
  {Gezari}, {Miller-Jones}, {Frederick}, {Sfaradi}, {Bietenholz}, {Horesh},
  {Fender}, {Garrappa}, {Ahumada}, {Andreoni}, {Belicki}, {Bellm},
  {B{\"o}ttcher}, {Brinnel}, {Burruss}, {Cenko}, {Coughlin}, {Cunningham},
  {Drake}, {Farrar}, {Feeney}, {Foley}, {Gal-Yam}, {Golkhou}, {Goobar},
  {Graham}, {Hammerstein}, {Helou}, {Hung}, {Kasliwal}, {Kilpatrick}, {Kong},
  {Kupfer}, {Laher}, {Mahabal}, {Masci}, {Necker}, {Nordin}, {Perley},
  {Rigault}, {Reusch}, {Rodriguez}, {Rojas-Bravo}, {Rusholme}, {Shupe},
  {Singer}, {Sollerman}, {Soumagnac}, {Stern}, {Taggart}, {van Santen}, {Ward},
  {Woudt}, \& {Yao}}]{2021NatAs...5..510S}
{Stein}, R., {Velzen}, S.~v., {Kowalski}, M., {et~al.} 2021, Nature Astronomy,
  5, 510

\bibitem[{{Tanaka} {et~al.}(2017){Tanaka}, {Buson}, \&
  {Kocevski}}]{2017ATel10791....1T}
{Tanaka}, Y.~T., {Buson}, S., \& {Kocevski}, D. 2017, The Astronomer's
  Telegram, 10791, 1

\bibitem[{{The IceCube-Gen2 Collaboration} {et~al.}(2020){The IceCube-Gen2
  Collaboration}, {:}, {Aartsen}, {Abbasi}, {Ackermann}, {Adams}, {Aguilar},
  {Ahlers}, {Ahrens}, {Alispach}, {Allison}, {Amin}, {Andeen}, {Anderson},
  {Ansseau}, {Anton}, {Arg{\"u}elles}, {Arlen}, {Auffenberg}, {Axani},
  {Bagherpour}, {Bai}, {Balagopal V.}, {Barbano}, {Bartos}, {Bastian}, {Basu},
  {Baum}, {Baur}, {Bay}, {Beatty}, {Becker}, {Becker Tjus}, {BenZvi}, {Berley},
  {Bernardini}, {Besson}, {Binder}, {Bindig}, {Blaufuss}, {Blot}, {Bohm},
  {Bohmer}, {B{\"o}ser}, {Botner}, {B{\"o}ttcher}, {Bourbeau}, {Bourbeau},
  {Bradascio}, {Braun}, {Bron}, {Brostean-Kaiser}, {Burgman}, {Burley},
  {Buscher}, {Busse}, {Bustamante}, {Campana}, {Carnie-Bronca}, {Carver},
  {Chen}, {Chen}, {Cheung}, {Chirkin}, {Choi}, {Clark}, {Clark}, {Classen},
  {Coleman}, {Collin}, {Connolly}, {Conrad}, {Coppin}, {Correa}, {Cowen},
  {Cross}, {Dave}, {Deaconu}, {De Clercq}, {DeLaunay}, {De Kockere},
  {Dembinski}, {Deoskar}, {De Ridder}, {Desai}, {Desiati}, {de Vries}, {de
  Wasseige}, {de With}, {DeYoung}, {Dharani}, {Diaz}, {D{\'\i}az-V{\'e}lez},
  {Dujmovic}, {Dunkman}, {DuVernois}, {Dvorak}, {Ehrhardt}, {Eller}, {Engel},
  {Evans}, {Evenson}, {Fahey}, {Farrag}, {Fazely}, {Felde}, {Fienberg},
  {Filimonov}, {Finley}, {Fischer}, {Fox}, {Franckowiak}, {Friedman}, {Fritz},
  {Gaisser}, {Gallagher}, {Ganster}, {Garcia-Fernand ez}, {Garrappa},
  {Gartner}, {Gerhardt}, {Gernhaeuser}, {Ghadimi}, {Glaser}, {Glauch},
  {Gl{\"u}senkamp}, {Goldschmidt}, {Gonzalez}, {Goswami}, {Grant},
  {Gr{\'e}goire}, {Griffith}, {Griswold}, {G{\"u}nd{\"u}z}, {Haack},
  {Hallgren}, {Halliday}, {Halve}, {Halzen}, {Hanson}, {Hanson}, {Hardin},
  {Haugen}, {Haungs}, {Hauser}, {Hebecker}, {Heinen}, {Heix}, {Helbing},
  {Hellauer}, {Henningsen}, {Hickford}, {Hignight}, {Hill}, {Hill}, {Hoffman},
  {Hoffmann}, {Hoffmann}, {Hoinka}, {Hokanson-Fasig}, {Holzapfel}, {Hoshina},
  {Huang}, {Huber}, {Huber}, {Huege}, {Hughes}, {Hultqvist}, {H{\"u}nnefeld},
  {Hussain}, {In}, {Iovine}, {Ishihara}, {Jansson}, {Japaridze}, {Jeong},
  {Jones}, {Jonske}, {Joppe}, {Kalekin}, {Kang}, {Kang}, {Kang}, {Kappes},
  {Kappesser}, {Karg}, {Karl}, {Karle}, {Katori}, {Katz}, {Kauer}, {Keivani},
  {Kellermann}, {Kelley}, {Kheirand ish}, {Kim}, {Kin}, {Kintscher}, {Kiryluk},
  {Kittler}, {Kleifges}, {Klein}, {Koirala}, {Kolanoski}, {K{\"o}pke},
  {Kopper}, {Kopper}, {Koskinen}, {Koundal}, {Kovacevich}, {Kowalski},
  {Krauss}, {Krings}, {Kr{\"u}ckl}, {Kulacz}, {Kurahashi}, {Lagunas Gualda},
  {Lahmann}, {Lanfranchi}, {Larson}, {Latif}, {Lauber}, {Lazar}, {Leonard},
  {Leszczy{\'n}ska}, {Li}, {Liu}, {Lohfink}, {LoSecco}, {Lozano Mariscal},
  {Lu}, {Lucarelli}, {Ludwig}, {L{\"u}nemann}, {Luszczak}, {Lyu}, {Ma},
  {Madsen}, {Maggi}, {Mahn}, {Makino}, {Mallik}, {Mancina}, {Mandalia},
  {Mari{\textcommabelow s}}, {Marka}, {Marka}, {Maruyama}, {Mase}, {Maunu},
  {McNally}, {Meagher}, {Medina}, {Meier}, {Meighen-Berger}, {Merz}, {Meyers},
  {Micallef}, {Mockler}, {Moment{\'e}}, {Montaruli}, {Moore}, {Morse},
  {Moulai}, {Muth}, {Naab}, {Nagai}, {Nam}, {Naumann}, {Necker}, {Neer},
  {Nelles}, {Nguy{\^e}n}, {Niederhausen}, {Nisa}, {Nowicki}, {Nygren},
  {Oberla}, {Obertacke Pollmann}, {Oehler}, {Olivas}, {O'Sullivan}, {Pan},
  {Pand ya}, {Pankova}, {Papp}, {Park}, {Parker}, {Paudel}, {Peiffer},
  {P{\'e}rez de los Heros}, {Petersen}, {Philippen}, {Pieloth}, {Pieper},
  {Pinfold}, {Pizzuto}, {Plaisier}, {Plum}, {Popovych}, {Porcelli}, {Prado
  Rodriguez}, {Price}, {Przybylski}, {Raab}, {Raissi}, {Rameez}, {Rauch},
  {Rawlins}, {Rea}, {Rehman}, {Reimann}, {Renschler}, {Renzi}, {Resconi},
  {Reusch}, {Rhode}, {Richman}, {Riedel}, {Riegel}, {Roberts}, {Robertson},
  {Roellinghoff}, {Rongen}, {Rott}, {Ruhe}, {Ryckbosch}, {Rysewyk Cantu},
  {Safa}, {Sanchez Herrera}, {Sand rock}, {Sandroos}, {Sandstrom}, {Santander},
  {Sarkar}, {Sarkar}, {Satalecka}, {Scharf}, {Schaufel}, {Schieler},
  {Schlunder}, {Schmidt}, {Schneider}, {Schneider}, {Schr{\"o}der},
  {Schumacher}, {Sclafani}, {Seckel}, {Seunarine}, {Shaevitz}, {Sharma},
  {Shefali}, {Silva}, {Smith}, {Smithers}, {Snihur}, {Soedingrekso}, {Soldin},
  {S{\"o}ldner-Rembold}, {Song}, {Southall}, {Spiczak}, {Spiering},
  {Stachurska}, {Stamatikos}, {Stanev}, {Stein}, {Stettner}, {Steuer},
  {Stezelberger}, {Stokstad}, {Strotjohann}, {St{\"u}rwald}, {Stuttard},
  {Sullivan}, {Taboada}, {Taketa}, {Tanaka}, {Tenholt}, {Ter-Antonyan},
  {Terliuk}, {Tilav}, {Tollefson}, {Tomankova}, {T{\"o}nnis}, {Torres},
  {Toscano}, {Tosi}, {Trettin}, {Tselengidou}, {Tung}, {Turcati}, {Turcotte},
  {Turley}, {Twagirayezu}, {Ty}, {Unger}, {Unland Elorrieta}, {Vand enbroucke},
  {van Eijk}, {van Eijndhoven}, {Vannerom}, {van Santen}, {Veberic},
  {Verpoest}, {Vieregg}, {Vraeghe}, {Walck}, {Watson}, {Weaver}, {Weindl},
  {Weinstock}, {Weiss}, {Weldert}, {Welling}, {Wendt}, {Werthebach},
  {Whitehorn}, {Wiebe}, {Wiebusch}, {Williams}, {Wissel}, {Wolf}, {Wood},
  {Woschnagg}, {Wrede}, {Wren}, {Wulff}, {Xu}, {Xu}, {Yanez}, {Yoshida},
  {Yuan}, {Zhang}, {Zierke}, \& {Z{\"o}cklein}}]{2020arXiv200804323T}
{The IceCube-Gen2 Collaboration}, {:}, {Aartsen}, M.~G., {et~al.} 2020,
  e-print, arXiv:2008.04323

\bibitem[{{Timmer} \& {Koenig}(1995)}]{Timmer1995}
{Timmer}, J. \& {Koenig}, M. 1995, \aap, 300, 707

\bibitem[{{van Velzen} {et~al.}(2021){van Velzen}, {Stein}, {Gilfanov},
  {Kowalski}, {Hayasaki}, {Reusch}, {Yao}, {Garrappa}, {Franckowiak}, {Gezari},
  {Nordin}, {Fremling}, {Sharma}, {Yan}, {Kool}, {Sollerman}, {Medvedev},
  {Sunyaev}, {Bellm}, {Dekany}, {Duev}, {Graham}, {Kasliwal}, {Laher},
  {Riddle}, \& {Rusholme}}]{vanvelzen2021}
{van Velzen}, S., {Stein}, R., {Gilfanov}, M., {et~al.} 2021, arXiv e-prints,
  arXiv:2111.09391

\bibitem[{{Zhang} {et~al.}(2019){Zhang}, {Fang}, {Li}, {Giannios},
  {B{\"o}ttcher}, \& {Buson}}]{2019ApJ...876..109Z}
{Zhang}, H., {Fang}, K., {Li}, H., {et~al.} 2019, \apj, 876, 109

\bibitem[{{Zhou} {et~al.}(2021){Zhou}, {Kamionkowski}, \& {Liang}}]{Zhou2021}
{Zhou}, B., {Kamionkowski}, M., \& {Liang}, Y.-f. 2021, \prd, 103, 123018

\end{thebibliography}

\onecolumn
\begin{table}
\centering
\caption{Percentage of 1000 simulations meeting the specific significance level in a given test.} \label{table:pvals}
\begin{tabular}{lclcccccc}
\hline\hline 
Selected flares & Signalness & $N_\mathrm{src}$ & \multicolumn{2}{c}{mean A.I.} &  \multicolumn{2}{c}{flaring sources}& \multicolumn{2}{c}{mean flux density} \\
 {}   &{(Y/N)} &{} &{$p\leq3\sigma$} &{$p\leq2\sigma$} &{$p\leq3\sigma$} &{$p\leq2\sigma$} &{$p\leq3\sigma$} &{$p\leq2\sigma$} \\
{(1)}   &{(2)} & {(3)} &{(4)} &{(5)} &{(6)} &{(7)} &{(8)} &{(9)}\\
\hline\hline 
 Any  & N & 1158 & 6.3 & 33.1 & 0.3 & 16.3 & 0 & 0.2\\
 $\geq$ Median & N &  1158 & 46.8 & 88.1 & 9.2 & 57.1 & 0 & 0.8 \\
 Maximum & N & 1158 & 100 & 100 & 70.9 & 100 & 0 & 0.9 \\
 Median & Y & 1158 & 15 & 50.7 & 1.7 & 31.1 & 0 & 1.6 \\
 Maximum & Y & 1158 & 41.5 & 82.3 & 10.8 & 67.9 & 0 & 1.4 \\
 Median, no flux cut & Y & 5000 & 0.1 & 40.2 & 0.1 & 13.5 & 0 & 0 \\
 Maximum, no flux cut & Y & 5000 & 0.5 & 97.8 & 0.9 & 60.7 & 0 & 0 \\
 Median, $\geq0.5$ Jy & Y & 5000 & 4.6 & 56 & 7.1 & 37.1 & 100 & 100 \\
 Maximum, $\geq0.5$ Jy & Y & 5000 & 60.4 & 98.9 & 43.0 & 82.9 & 100 & 100 \\
 Median, $\geq1.0$ Jy & Y & 5000 & 4.0 & 84.1 & 40.1 & 78.7 & 100 & 100 \\
 Maximum, $\geq1.0$ Jy & Y & 5000 & 42.8 & 100 & 94.9 & 100 & 100 & 100 \\
 \end{tabular}
\tablefoot{Col (1) indicates which flares were selected to be associated with neutrinos and in the case of 5000 simulations also which flux density cutoff was used to pick the associated sources. Col (2) indicates whether the signalness of the neutrinos was accounted for in the simulations. Col (3) gives the number of sources in the simulation where 1158 mimics the real CGRaBS sample of H21 and 5000 the new OVRO monitoring sample. Cols (4) and (5) give the fraction of simulations (in $\%$) that have smaller post-trial p-value than $3\sigma$ (col 4) or $2\sigma$ (col 5) when looking at the mean activity index of the associated samples. Cols (6) and (7) give the same information for the test with the number of flaring sources, and Cols (8) and (9) for the mean flux density of the associated sample (see text for details). The significance for the CGRaBS sample in H21 (without accounting for the trial factor due to multiple studied samples) was $2.3\sigma$ (p-value=0.01) for the mean A.I., $2.8\sigma$ (p-value=0.005) for the number of flaring sources, and $2.0\sigma$ (p-value=0.027) for the mean flux density.}
\end{table}
\twocolumn

\end{document}